\documentclass[12pt]{article}
\usepackage{graphicx}

\begin{document}
\pagenumbering{arabic}
\begin{titlepage}
\title{The teleparallel equivalent of general relativity }

\author{J. W. Maluf$^*$ \\
Instituto de F\'{\i}sica \\
Universidade de 
Bras\'{\i}lia \\
C.P. 04385 \\ 70.919-970 
Bras\'{\i}lia DF \\ Brazil}
\date{}
\maketitle
\bigskip

\begin{abstract}
A review of the teleparallel equivalent of general relativity is presented.
It is emphasized that general relativity may be formulated in terms of the 
tetrad fields and of the torsion tensor, and that this geometrical formulation
leads to alternative insights into the theory. The equivalence with the 
standard formulation in terms of the metric and curvature tensors takes place 
at the level of field equations. The review starts with a brief account of
the history of teleparallel theories of gravity. Then the ordinary 
interpretation of the tetrad fields as reference frames adapted to arbitrary 
observers in space-time is discussed, and the tensor of inertial accelerations
on frames is obtained. It is shown that the Lagrangian and Hamiltonian field 
equations allow to define the energy, momentum and angular momentum of the
gravitational field, as surface integrals of the field quantities. In the 
phase space of the theory, these quantities satisfy the algebra of the 
Poincar\'{e} group. 
\end{abstract}
\thispagestyle{empty}
\vfill

\bigskip
($^*$) wadih@unb.br, jwmaluf@gmail.com\par
\end{titlepage}

\section{Introduction}
The teleparallel equivalent of general relativity (TEGR) is an alternative 
geometrical formulation of Einstein's general relativity. It may be formulated
either in terms of the tetrad fields $e^a\,_\mu$ and of an independent 
SO(3,1) (Lorentz) connection $\omega_{\mu ab}$
\footnote{ $a,b,...$ are SO(3,1) or Lorentz indices, and 
$\mu, \nu, ...$ are space-time indices.},
or in terms of the tetrad fields only. The simplest realization, namely, a 
teleparallel theory constructed solely out of $e^a\,_\mu$, 
preserves the physical features of the theory.
Given a set of tetrad fields, it is 
possible to construct the metric tensor $g_{\mu\nu}$,
the Christoffel symbols $^0\Gamma^\lambda_{\mu\nu}$, and the torsion-free 
Levi-Civita connection $^0\omega_{\mu ab}(e)$, to be defined ahead. It is also 
possible to construct the Weitzenb\"{o}ck connection 
$\Gamma^\lambda_{\mu\nu}=e^{a\lambda} \partial_\mu e_{a\nu}$ \cite{Weit}. The
curvature tensor constructed out of the latter vanishes identically. In 
the realm of a theory constructed out of the tetrad fields only, it is possible
to address geometrical issues of both the Weitzenb\"{o}ck and Riemannian
geometries. Therefore, the tetrad theory of gravity is a geometrical framework
more general than (and consistent with) the Riemannian geometry.

Teleparallelism is a geometrical setting where it is possible to establish the
notion of distant parallelism. For this purpose, one has to fix a particular 
frame, but in the TEGR any frame is allowed in view of the field
equations. In a space-time endowed with a set of tetrad 
fields, two  vectors at distant points are called parallel \cite{Mol} if they 
have identical components with respect to the local tetrads at the points 
considered. Thus, consider a vector field $V^\mu(x)$. At the point $x^\lambda$
its tetrad components are $V^a(x)=e^a\,_\mu(x)V^\mu(x)$. For the 
tetrad components $V^a(x+dx)$ at $x^\lambda + dx^\lambda$, it is easy to see 
that $V^a(x+dx)=V^a(x)+\nabla V^a(x)$, where 
$\nabla V^a(x)=e^a\,_\mu(\nabla_\lambda V^\mu)dx^\lambda$. The covariant 
derivative $\nabla$ is constructed out of the Weitzenb\"{o}ck connection. 
Therefore, the 
vanishing of this covariant derivative defines a condition for absolute 
parallelism in space-time. Since $\nabla_\lambda e^a\,_\mu\equiv 0$, the
tetrad fields constitute a set of auto-parallel fields. The covariant 
derivative is not covariant under local SO(3,1) (Lorentz) transformations.
Geometrical quantities invariant under local Lorentz transformations can be
freely rotated in every point of the space-time, and for such quantities it is
not natural to establish the idea of distant parallelism. The lack of local
SO(3,1) symmetry does not mean that a particular frame is distinguished. All
physical frames are solutions of the field equations. The teleparallel
geometry may be understood as a limiting case of the more general 
Riemann-Cartan geometries \cite{Hehl1,Hehl2}, which are defined by 
arbitrary configurations of the curvature and torsion tensors. 

The most simple geometrical quantities that are obtained from the tetrad fields
are the metric tensor $g_{\mu\nu}=e^a\,_\mu e^b\,_\nu\eta_{ab}$, where 
$\eta_{ab}$ is the flat space-time metric tensor, and 
the torsion tensor $T_{a\mu\nu}=\partial_\mu e_{a\nu}-\partial_\nu e_{a\mu}$. 
The tensor $T_{\lambda \mu\nu}=e^a\,_\lambda T_{a\mu\nu}$ is precisely the 
torsion of the Weitzenb\"{o}ck connection. Out of 
$T_{abc}=e_b\,^\mu e_c\,^\nu T_{a\mu\nu}$ one may construct the three 
Weitzenb\"{o}ck invariants:
$I_1=A\,T^{abc}T_{abc}$, $I_2=B\,T^{abc}T_{bac}$ and $I_3=C\,T^aT_a$, where 
$T_a=T^b\,_{ba}$, and  $A$, $B$, $C$ are arbitrary numerical 
constants. Arbitrary values of the
constants $A$, $B$ and $C$ lead to arbitrary teleparallel theories of gravity,
defined by the Lagrangian density $L=e(AI_1+BI_2+CI_3)$, where 
$e=\det(e^a\,_\mu)$. In the period 1928-31 Einstein became interested in
teleparallel theories as a possible framework for unification.
In 1929 Einstein noted
that the field equations obtained from the theory for which $A=1/4$, $B=1/2$
and $C=-1$ are symmetric in the two free space-time indices, and that the 
resulting linearised theory describes the weak gravitational 
field. He allowed the three constants to acquire values slightly different
from the values above, and pursued the formulation of a unified
field theory of gravitation and electromagnetism. The extra 6 of the 
16 degrees of freedom of the tetrad field would be identified
with the electromagnetic fields. Cornelius Lanczos noted that
the invariant defined by $A=1/4$, $B=1/2$ and 
$C=-1$ is essentially equivalent to the Riemannian scalar curvature $R$, up
to a total divergence. These facts are reported in the historical 
account by T. Sauer \cite{Sauer}. Einstein did not succeed in arriving at a
faithful and consistent tensor-like description of the electromagnetic field 
equations in this approach. One of the difficulties of the unification 
program was the large freedom in the choice of the field equations. It was 
not possible to justify an uniquely determined set of acceptable equations, 
and for this reason Einstein abandoned the approach \cite{Sauer}. In this 
review we argue that the extra 6 degrees of freedom of the tetrad fields are
taken to fix the reference frame in space-time. At the level of 
Hamiltonian field equations, they lead to 6 primary, first class constraints,
and also to the definition of the gravitational angular momentum.

Teleparallel gravity was reconsidered in 1976 by Cho \cite{Cho,Cho2},
who derived a tetrad theory of gravity as a gauge theory of the 
translation group, although the theory was not described in the geometrical 
framework of teleparallelism. Cho argued that the resulting 
Einstein-Cartan type theory is the unique gauge theory of the Poincar\'{e} 
group $P_4$, if the Lagrangian density is constructed out of the lowest 
possible combinations of the field strengths \cite{Cho2}. At about the same 
time, teleparallel theories were investigated as gravity theories in
the Weitzenb\"{o}ck space-time. The motivation for this renewed interest was 
the analysis by Hayashi \cite{Hayashi} in 1977 on the gauge theory of 
the translation group in connection with the space-time torsion. 
Hayashi observed that a gravitational theory based on the Weitzenb\"{o}ck 
space-time may be interpreted as a gauge theory of the translation group, 
where the gauge field is identified as a part of the tetrad fields (Cho
\cite{Cho2} made the same identification earlier in 1976). 
However, no specific theory
was analysed by Hayashi. In 1979, Hayashi and Shirafuji \cite{Hay} investigated
in detail a general class of teleparallel theories. The theory was
called ``New General Relativity'', since it was a reconsideration of
Einstein's previous approach.
They again concluded that for a certain fixation of the constant parameters, 
the Lagrangian density reduces to the scalar curvature 
density $eR$ of the Riemannian geometry. They established a one-parameter 
theory that deviates from the standard formulation of general relativity.
In the same period, Hehl \cite{Hehl} and Nitsch \cite{Nitsch} 
addressed a general class of gravity theories in the Riemann-Cartan geometry, 
the ``Poincar\'e Gauge Theory of Gravity'', with the purpose of investigating
the Yang-Mills-type structure of the field equations of gravity.
These are theories with {\it a priori} independent connection and tetrad 
fields, which include teleparallel theories as particular cases, 
and one of these theories is equivalent to the standard general
relativity. 

Metric-affine theories of gravity are theories constructed out of a set
of tetrad fields $e^a\,_\mu$ (or a coframe one-form) and an arbitrary 
connection $\omega_{\mu ab}$. A metric-teleparallel theory belongs to a
particular class of metric-affine theories where the Lagrangian density is 
given by a suitable invariant
quadratic in the torsion tensor, constrained by the condition that the 
curvature tensor of the connection $\omega_{\mu ab}$ vanishes. One specific
theory is equivalent to Einstein's general relativity in the sense that the 
field equations for the tetrad fields (or metric tensor) are precisely 
Einstein's equations. We will not address this formulation in the present
review, because the connection $\omega_{\mu ab}$ introduces an additional 
geometric structure. In the context of the TEGR, this connection plays no 
role in the dynamics of the tetrad fields, and consequently in the space-time
geometry (see Sections 3.1 and 7).

The metric-teleparallel theory equivalent to the standard general relativity
was critically analysed by Kopczy\'nski \cite{Kop}, who concluded that the
teleparallel field equations do not give full information about the 
teleparallel connection $\omega_{\mu ab}$, and leads to a non-predictable 
behaviour of torsion. Nester \cite{Nester1} addressed the difficulties raised
by Kopczy\'nski, and found that they are not generic, but for certain
special solutions there is a problematic gauge freedom. Nester also
addressed the canonical analysis of the TEGR, with the purpose of obtaining
a new proof of the positivity of the gravitational energy \cite{Nester2}.
In the context of the teleparallel theory
equivalent to the standard general relativity, Mielke \cite{Mielke} 
investigated a theory formulated in terms of Ashtekar's complex variables.
In this approach, the field equations acquire a Yang-Mills-type structure
with respect to a self-dual connection.

The Hamiltonian formulation of the TEGR was investigated in 1994 in 
Ref. \cite{Maluf1}. In order to simplify the analysis, the canonical 3+1 
decomposition and the constraint algebra were carried out under the imposition
of Schwinger's time gauge condition \cite{Schwinger}. 
The advantage of taking into account this gauge condition is that the 
resulting canonical structure and constraint algebra is
structurally similar to the ADM Hamiltonian formulation \cite{ADM} of the 
standard general relativity. This analysis was possible because the 
Lagrangian density and the field equations were written in a compact form, in
terms of the tensor $\Sigma^{abc}$, which sometimes is called superpotential,
and which will be defined in Section 3.
The emergence of a scalar density as a total divergence of the trace of the 
torsion tensor, in the Hamiltonian constraint
of the theory, motivated the interpretation of this term as the gravitational
energy density. The integral of this term over the whole three-dimensional
space yields the ADM energy, for suitable asymptotic boundary conditions, 
and a first covariant expression for the gravitational 
energy, in the realm of the TEGR, was presented in Ref. \cite{Maluf2}. 
The torsion tensor cannot be made to vanish at a point in 
space-time by means of a coordinate transformation. Therefore, criticisms
based on the principle of equivalence, which rest on the reduction of the
metric tensor to the Minkowski metric tensor at any point in space-time by
means of a coordinate transformation, do not apply to the definition of
gravitational energy that arises in the TEGR. In the framework of the metrical
formulation of general relativity it is not possible to construct any 
non-trivial scalar density that depends on the second order derivatives of 
the metric tensor, that could be interpreted as the gravitational 
energy-momentum density. It is known that all gravitational energy-momentum 
pseudo-tensors depend on quantities that are badly behaved under coordinate 
transformations, since they depend on the coordinate system.

An expression for the gravitational energy-momentum in the TEGR as a surface
integral, without the imposition of Schwinger's time gauge condition, was 
first presented in 1999 in Ref. \cite{Maluf3}, 
and investigated in detail in Ref. \cite{Maluf4}. The full 
Hamiltonian formulation, together with the constraint algebra, was 
analysed in Ref. \cite{Maluf5}, and further refined in Ref. \cite{Maluf6}.
The gravitational energy-momentum vector $P^a$ satisfies continuity (or 
balance) equations \cite{Maluf7}, which lead to conservation laws for $P^a$ 
and to a definition of the gravitational energy-momentum tensor. These issues
will presented in detail in this review.

Teleparallel gravity has been investigated by Aldrovandi and Pereira, as a
gauge theory of the translation group. In similarity to Hayashi's approach
\cite{Hayashi}, they identify the gravitational potential as a non-trivial
part of the tetrad field, and gravity is described in the Weitzenb\"ock
space-time. Their approach is presented in Refs. 
\cite{Aldrovandi1,Aldrovandi2}. Teleparallel gravity has been readdressed by
Obukhov and Pereira \cite{OP} in the geometrical framework of metric-affine 
theories, and further reconsidered by Obukhov and 
Rubilar \cite{Obukhov1,Obukhov2}, with the purpose of investigating the 
transformation (covariance) properties and conserved currents in tetrad 
theories of gravity. They analysed the problem of consistently defining the
gravitational energy-momentum and, in particular, the problem of 
regularization of the expression of the gravitational energy-momentum (see
also Ref. \cite{Itin}). This issue will also be addressed later on in the 
present geometrical framework.

This review aims at summarizing the work that has been developed since 1994
in the establishment of the TEGR, emphasizing the 
crucial role of tetrad fields as frames adapted to arbitrary
observers in space-time. Accelerated frames are frames with torsion 
\cite{Schucking}. The tetrad fields describe at the same time the 
gravitational field and the frame. In particular, the torsion tensor 
$T_{a\mu\nu}$ plays an important role in the definition of the tensor of 
inertial accelerations on frames, a quantity that evidently is frame
dependent. It is natural to consider the TEGR 
as an alternative description of
the gravitational field, because the theory is constructed out of 
$T_{a\mu\nu}$.  We will argue that the frame dependence of quantities
such as the gravitational energy-momentum vector is a physically consistent
feature, since the concepts that are valid in the special theory of
relativity are also valid in the general theory. There is no clear cut 
division of the physical concepts in the special and general theories of 
relativity. The introduction of the gravitational field does not modify the
frame dependence of the energy of a particle in special relativity (which is 
the zero component of a vector), and therefore the gravitational energy of a 
black hole, for instance, viewed as a particle at very large
distances, should also be frame dependent. We will also briefly
review the Hamiltonian formulation of the TEGR, which is of fundamental
importance for a complete understanding of the theory. \par
\bigskip

\noindent {\bf Notation}:
space-time indices $\mu, \nu, ...$ and SO(3,1) (Lorentz) indices
$a, b, ...$ run from 0 to 3. Time and space indices are indicated according to
$\mu=0,i,\;\;a=(0),(i)$. The tetrad fields are represented by $e^a\,_\mu$, and 
the torsion tensor by 
$T_{a\mu\nu}=\partial_\mu e_{a\nu}-\partial_\nu e_{a\mu}$.
The flat, tangent space Minkowski space-time metric tensor raises and lowers 
tetrad indices, and is fixed by 
$\eta_{ab}= e_{a\mu} e_{b\nu}g^{\mu\nu}=(-1,+1,+1,+1)$.
The frame components are given by the inverse tetrads 
$\lbrace e_a\,^\mu \rbrace$, although we may as well refer to 
$\lbrace e^a\,_\mu\rbrace$ as the frame. The determinant of the tetrad field is 
represented by $e=\det(e^a\,_\mu)$.

The torsion tensor defined above is often related to the object of
anholonomity $\Omega^\lambda\,_{\mu\nu}$ via 
$\Omega^\lambda\,_{\mu\nu}= e_a\,^\lambda T^a\,_{\mu\nu}$.
However, we assume that the space-time geometry is defined by the 
tetrad fields only, and in this case the only
possible non-trivial definition for the torsion tensor is given by
$T^a\,_{\mu\nu}$. This torsion tensor is related to the 
antisymmetric part of the Weitzenb\"ock  connection 
$\Gamma^\lambda_{\mu\nu}=e^{a\lambda}\partial_\mu e_{a\nu}$, which
establishes the Weitzenb\"ock space-time. The metric and torsion-free 
Christoffel symbols are denoted by $^0\Gamma^\lambda_{\mu\nu}$, and the
associated torsion-free Levi-Civita connection $^0\omega_{\mu ab}$ is defined 
by Eq. (\ref{1-7}). These connections are related by Eq. (\ref{2-1-8}) below.

\section{The tetrad fields and reference frames}

A set of tetrad fields is defined  by four orthonormal, linearly independent 
vector fields in space-time,
$\lbrace e^{(0)}\,_\mu, e^{(1)}\,_\mu, e^{(2)}\,_\mu, e^{(3)}\,_\mu\rbrace$,
which establish the local reference frame of an observer that moves along a 
trajectory $C$, represented by the worldline $x^\mu(\tau)$ 
\cite{Hehl3,Maluf8,Maluf9} ($\tau$ is the proper time of the observer). 
The components $e^{(0)}\,_\mu$ and $e^{(i)}\,_\mu$ are timelike and spacelike
vectors, respectively; $e^a\,_\mu$ transforms as covariant vector fields under
coordinate transformations, and as contravariant vector fields under SO(3,1)
(Lorentz) transformations, i.e., $\tilde{e}^a\,_\mu=\Lambda^a\,_b\,e^b\,_\mu$,
where the matrices $\lbrace \Lambda^a\,_b\rbrace$ are representations of the
SO(3,1) group and satisfy $\Lambda^a\,_c\Lambda^b\,_d\, \eta_{ab}=\eta_{cd}$.
The metric tensor $g_{\mu\nu}$ is obtained by 
the relation $e^a\,_\mu e^b\,_\nu \eta_{ab}=g_{\mu\nu}$.
The tetrad fields $e^a\,_\mu$ allow the projection of vectors and tensors in 
space-time in the local frame of an observer. 
In order to measure field quantities with magnitude and direction, an observer
must project these quantities on the frame 
carried by the observer. The projection of a vector $V^{\mu}(x)$ at a position
$x^\mu$, on a
particular frame, is determined by $V^{a}(x) = e^{a}\,_{\mu}(x)\,V^{\mu}(x)$.

Given a  worldline $C$ of an observer, represented by 
$x^\mu(\tau)$, the velocity of the observer along $C$ is denoted by 
$u^\mu(\tau)=dx^\mu/d\tau$. We identify the observer's
velocity with the $a=(0)$ component of $e_a\,^\mu$. Thus,
$e_{(0)}\,^\mu=u^\mu(\tau)/c$. The acceleration $a^\mu$ of the observer
is given by the absolute derivative of $u^\mu$ along $C$, 

\begin{equation}
a^\mu= {{Du^\mu}\over{d\tau}} =c{{De_{(0)}\,^\mu}\over {d\tau}} =
c\,u^\alpha \nabla_\alpha e_{(0)}\,^\mu\,, 
\label{1-1}
\end{equation}
where the covariant derivative is constructed out of the Christoffel symbols
$^0\Gamma^\mu_{\alpha\beta}$. The last equality follows from

\begin{eqnarray}
{{De_{(0)}\,^\mu}\over {d\tau}}&=& {{de_{(0)}\,^\mu}\over {d\tau}}+\,\,
^0\Gamma^\mu_{\alpha\beta}\, {{dx^\alpha}\over{d \tau}}\, e_{(0)}\,^\beta
\nonumber \\
&=& {{dx^\alpha} \over {d\tau}}\,
{{\partial e_{(0)}\,^\mu}\over {\partial x^\alpha}} +\,\,
^0\Gamma^\mu_{\alpha\beta}\, {{dx^\alpha}\over{d \tau}}\, e_{(0)}\,^\beta
\nonumber  \\
&=& u^\alpha \,\nabla _\alpha e_{(0)}\,^\mu\,.
\label{1-2}
\end{eqnarray}

Thus, $e_a\,^\mu$ yields the velocity and acceleration of an observer along 
the worldline. Therefore, a given set of tetrad fields, for which 
$e_{(0)}\,^\mu$ describes a congruence of timelike curves, is adapted to a 
particular class of observers, namely, to observers characterized by the 
velocity field $u^\mu=c\,e_{(0)}\,^\mu$, endowed with acceleration $a^\mu$. If
$e^a\,_\mu \rightarrow \delta^a_\mu$ in the limit 
$r \rightarrow \infty$, then $e^a\,_\mu$ is adapted to static
observers at spacelike infinity. 

The geometrical characterization of tetrad fields as an observer's 
frame may be given by considering the acceleration of the frame along an 
arbitrary path $x^\mu(\tau)$ of the observer. The acceleration 
of the whole frame is determined by the absolute derivative of $e_a\,^\mu$ 
along $x^\mu(\tau)$. Thus, assuming that the observer carries an orthonormal 
tetrad  frame $e_a\,^\mu$, the acceleration of the frame along the path is 
given by \cite{Mashh2,Mashh3}

\begin{equation}
{{D e_a\,^\mu} \over {d\tau}}=\phi_a\,^b\,e_b\,^\mu\,,
\label{1-3}
\end{equation}
where $\phi_{ab}$ is the antisymmetric acceleration tensor. According to Refs. 
\cite{Mashh2,Mashh3}, in analogy with the Faraday tensor we can identify 
$\phi_{ab} \rightarrow ({\bf a}/c, {\bf \Omega})$, where ${\bf a}$ is the 
translational acceleration ($\phi_{(0)(i)}=a_{(i)}/c$) and ${\bf \Omega}$ is the 
frequency of rotation of the local spatial frame  with respect to a 
non-rotating, Fermi-Walker transported frame. 
It follows from Eq. (\ref{1-3}) that

\begin{equation}
\phi_a\,^b= e^b\,_\mu {{D e_a\,^\mu} \over {d\tau}}=
e^b\,_\mu \,u^\lambda\nabla_\lambda e_a\,^\mu\,.
\label{1-4}
\end{equation}

The acceleration vector $a^\mu$ defined by Eq. (\ref{1-1}) may be projected
on a frame in order to yield

\begin{equation}
a^b= e^b\,_\mu a^\mu=c\,e^b\,_\mu u^\alpha \nabla_\alpha
e_{(0)}\,^\mu=c\,\phi_{(0)}\,^b\,.
\label{1-5}
\end{equation}
Thus, $a^\mu$ and $\phi_{(0)(i)}$ are not different translational 
accelerations of the frame. The expression of $a^\mu$ given by Eq. (\ref{1-1}) 
may be rewritten as

\begin{eqnarray}
a^\mu/c&=& u^\alpha \nabla_\alpha e_{(0)}\,^\mu 
=u^\alpha \nabla_\alpha u^\mu =
{{dx^\alpha}\over {ds}}\biggl(
{{\partial u^\mu}\over{\partial x^\alpha}}
+\,\,^0\Gamma^\mu_{\alpha\beta}u^\beta \biggr) \nonumber \\
&=&{{d^2 x^\mu}\over {ds^2}}+\,\,^0\Gamma^\mu_{\alpha\beta}
{{dx^\alpha}\over{ds}} {{dx^\beta}\over{ds}}\,,
\label{1-6}
\end{eqnarray}
where $\,\,^0\Gamma^\mu_{\alpha\beta}$ are the Christoffel symbols.
We see that if $u^\mu=c\,e_{(0)}\,^\mu$ represents a geodesic
trajectory, then the frame is in free fall and 
$a^\mu/c=\phi_{(0)(i)}=0$. Therefore we conclude that non-vanishing
values of the latter quantities do represent inertial accelerations
of the frame.

Since the tetrads are orthonormal vectors, we may  rewrite Eq. 
(\ref{1-4}) as
$\phi_a\,^b= -u^\lambda e_a\,^\mu \nabla_\lambda e^b\,_\mu$, 
where $\nabla_\lambda e^b\,_\mu=\partial_\lambda e^b\,_\mu-\,\,
^0\Gamma^\sigma_{\lambda \mu} e^b\,_\sigma$. Now we take into account
the identity 

$$\partial_\lambda e^b\,_\mu-\,\,
^0\Gamma^\sigma_{\lambda \mu} e^b\,_\sigma+\,\,
^0\omega_\lambda\,^b\,_c e^c\,_\mu=0\,,$$ 
where $^0\omega_\lambda\,^b\,_c$ is the metric compatible, torsion free
Levi-Civita connection, 

\begin{eqnarray}
^0\omega_{\mu ab}&=&-{1\over 2}e^c\,_\mu(
\Omega_{abc}-\Omega_{bac}-\Omega_{cab})\,,  \nonumber \\
\Omega_{abc}&=&e_{a\nu}(e_b\,^\mu\partial_\mu
e_c\,^\nu-e_c\,^\mu\partial_\mu e_b\,^\nu)\,,
\label{1-7}
\end{eqnarray}
and express $\phi_a\,^b$ as 
$\phi_a\,^b=c\,e_{(0)}\,^\mu(\,\,^0\omega_\mu\,^b\,_a)$. 
At last we consider the identity $\,\,^0\omega_\mu\,^a\,_b=
-K_\mu\,^a\,_b$, where $K_\mu\,^a\,_b$ is the 
contortion tensor defined by

\begin{equation}
K_{\mu ab}={1\over 2}e_a\,^\lambda e_b\,^\nu(T_{\lambda \mu\nu}+
T_{\nu\lambda\mu}+T_{\mu\lambda\nu})\,,
\label{1-9}
\end{equation}
and $T_{\lambda \mu\nu}=e^a\,_\lambda T_{a\mu\nu}$ (see 
Section 3.1 ahead or 
Eq. (4) of Ref. \cite{Maluf10}; the identity may be obtained by direct
calculation). After simple manipulations we finally obtain

\begin{equation}
\phi_{ab}={c\over 2} \lbrack T_{(0)ab}+T_{a(0)b}-T_{b(0)a}
\rbrack\,.
\label{1-10}
\end{equation}

The expression above is clearly not invariant under local SO(3,1)
transformations, but is invariant under coordinate transformations.
The values of $\phi_{ab}$ for given tetrad fields 
may be used to characterize the frame. We interpret $\phi_{ab}$ as the
inertial accelerations along the trajectory $x^\mu(\tau)$.

Therefore, given any set of tetrad fields for an arbitrary space-time, its 
geometrical interpretation may be obtained {\bf (i)} either by suitably 
identifying the velocity $u^\mu=c\,e_{(0)}\,^{\mu}$ of the field of 
observers, together with the orientation in the three-dimensional space of 
the components $e_{(1)}\,^\mu,e_{(2)}\,^\mu,e_{(3)}\,^\mu$, or {\bf (ii)} by 
the values of the 
acceleration tensor $\phi_{ab}=-\phi_{ba}$, which characterize
the inertial state of the frame. The condition $e_{(0)}\,^{\mu}=u^\mu/c$ fixes
only the three components $e_{(0)}\,^1$, $e_{(0)}\,^2$, $e_{(0)}\,^3$, because
the component $e_{(0)}\,^0$ is determined by the normalization condition 
$u^\mu u^\nu g_{\mu\nu}=-c^2$. In both cases, the fixation of the frame 
requires the fixation of 6 components of the tetrad fields.

Fermi-Walker transported frames define a standard of non-rotation for 
accelerated observers. These are frames for which the frequency of rotation
$\phi_{(i)(j)}$ vanishes \cite{Maluf9}. Suppose that a frame is given such 
that $\phi_{(j)(k)}\ne 0$. We may transform this frame into a Fermi-Walker
transported frame by means of the following procedure. First we note that 
in terms of the torsion tensor the quantities $\phi_{(j)(k)}$ are 
written as

\begin{equation}
\phi_{(i)(j)}={1\over 2} \lbrack
e_{(i)}\,^\mu e_{(j)}\,^\nu T_{(0)\mu\nu} +
e_{(0)}\,^\mu e_{(j)}\,^\nu T_{(i)\mu\nu} -
e_{(0)}\,^\mu e_{(i)}\,^\nu T_{(j)\mu\nu}\rbrack\,.
\label{1-2-17}
\end{equation}
Under a local Lorentz transformation of the spatial components we have 

\begin{eqnarray}
\tilde{e}_{(i)}\,^\mu &=& \Lambda_{(i)}\,^{(k)} e_{(k)}\,^\mu \,, 
\label{1-2-18} \\
\tilde{T}_{(i)\mu\nu}&=&
\partial_\mu \tilde{e}_{(i)\nu} -
\partial_\nu \tilde{e}_{(i)\mu} \nonumber \\
{}&=&\Lambda_{(i)}\,^{(k)} T_{(k)\mu\nu}+
\lbrack \partial_\mu \Lambda_{(i)}\,^{(k)}\rbrack e_{(k)\nu}-
\lbrack \partial_\nu \Lambda_{(i)}\,^{(k)}\rbrack e_{(k)\mu}\,.
\label{1-2-19}
\end{eqnarray}
The coefficients $\lbrace \Lambda_{(i)}\,^{(j)}(x) \rbrace$ of the spatial
components of the local Lorentz transformation are fixed by requiring
$\tilde{\phi}_{(i)(j)}=0$. It is possible to show that 
for given non-vanishing values of the quantities $\phi_{(j)(k)}$, the 
condition $\tilde{\phi}_{(i)(j)}=0$ is achieved provided the coefficients
$\lbrace \Lambda_{(i)}\,^{(j)} \rbrace$ of the Lorentz transformation 
satisfy the equation \cite{Maluf9}

\begin{equation}
e_{(0)}\,^\mu \Lambda^{(j)}\,_{(m)} \partial_\mu \Lambda_{(j)(k)}-
\phi_{(k)(m)}=0\,.
\label{1-2-20}
\end{equation}
Thus, given an arbitrary frame, it is possible, at least formally, to
rotate the frame and obtain a Fermi-Walker transported frame. We note that 
the local Lorentz transformation (\ref{1-2-18}) does not affect the component
$e_{(0)}\,^\mu$.

\section{The Lagrangian formulation of the TEGR}

Teleparallel theories of gravity are constructed out of the tetrad fields
$e_{a\mu}$ and the torsion tensor 
$T_{a\mu\nu}=\partial_\mu e_{a\nu}-\partial_\nu e_{a\mu}$. The class of 
theories whose field equations are second order differential equations are 
established from a Lagrangian density constructed out of the Weitzenb\"{o}ck
invariants $T^{abc}T_{abc}$, $T^{abc}T_{bac}$ 
and $T^aT_a$, where $T_a=T^b\,_{ba}$. The equivalence of
a particular teleparallel theory  with Einstein's general relativity is
verified by means of algebraic identities between the tetrad fields, the 
torsion tensor, the contorsion tensor and the SO(3,1) (Lorentz) connection
$^0\omega_{\mu ab}$, which is the torsion free, Levi-Civita connection. We 
will first present the identities, and then we discuss the field equations,
the balance equations and the energy-momentum tensor of the gravitational 
field.

\subsection{Geometrical identities}

The most important identity relates the Levi-Civita connection
$^0\omega_{\mu ab}$ given by Eq. (\ref{1-7}) with the contorsion tensor 
$K_{\mu ab}$, defined by

\begin{equation}
K_{\mu ab}={1\over 2}e_a\,^\lambda e_b\,^\nu(T_{\lambda \mu\nu}+
T_{\nu\lambda\mu}+T_{\mu\lambda\nu})\,.
\label{2-1-1}
\end{equation}
The identity reads

\begin{equation}
^0\omega_{\mu ab}=-K_{\mu ab}\,.
\label{2-1-2}
\end{equation}
This identity may be obtained by  
direct calculations, or by means of the following procedure. Let us consider
a four-dimensional pseudo-Riemannian manifold endowed with a set of tetrad 
fields $e^a\,_\mu$ and an independent, arbitrary
SO(3,1) connection $\omega_{\mu ab}$. These quantities define the torsion 
and curvature tensors according to

\begin{equation}
{\cal T}^a\,_{\mu \nu}(e,\omega)=
\partial_\mu e^a\,_\nu-\partial_\nu e^a\,_\mu
+\omega_\mu\,^a\,_b\,e^b\,_\nu- \omega_\nu\,^a\,_b\,e^b\,_\mu\;,
\label{2-1-3}
\end{equation}

\begin{equation}
R^a\,_{b\mu\nu}(\omega)=\partial_\mu \omega_\nu\,^a\,_b
-\partial_\nu \omega_\mu\,^a\,_b
+\omega_\mu\,^a\,_c\, \omega_\nu\,^c\,_b
-\omega_\nu\,^a\,_c\, \omega_\mu\,^c\,_b\,,
\label{2-1-4}
\end{equation}
respectively (our notation is the same as in Ref. \cite{Eguchi}).
The equation that defines ${\cal T}_{a \mu\nu}(e,\omega)$ can 
be solved for $\omega_{\mu ab}$. After a number manipulations, where we
take into account the antisymmetry $\omega_{\mu ab} =-\omega_{\mu ba}$,
it is possible to obtain the identity

\begin{equation}
\omega_{\mu ab}=\;^0\omega_{\mu ab}(e) + {\cal K}_{\mu ab}\,,
\label{2-1-5}
\end{equation}
where

\begin{equation}
{\cal K}_{\mu ab}=
{1\over 2}e_a\,^\lambda e_b\,^\nu({\cal T}_{\lambda \mu\nu}+
{\cal T}_{\nu\lambda\mu}+{\cal T}_{\mu\lambda\nu})\,.
\label{2-1-6}
\end{equation}
It is possible to verify that the arbitrary connection $\omega_{\mu ab}$ 
plays no role in the dynamics of tetrad  fields in the TEGR, as we conclude
from Eqs. (6) and (9) of Ref. \cite{Maluf1}. Therefore we dispense with this
connection, and require it to vanish: $\omega_{\mu ab}=0$. As a consequence, 
${\cal T}_{a\mu\nu}$ reduces to $T_{a\mu\nu}$, and Eq. (\ref{2-1-5}) 
to Eq. (\ref{2-1-2}).

The covariant derivative of the tetrad fields with respect to the Christoffel
symbols and the Levi-Civita connection $^0\omega_{\mu ab}$ is identically
vanishing,

\begin{equation}
D_\mu e^a\,_\nu =\partial_\mu e^a\,_\nu +\,\,^0\omega_\mu\,^a\,_ b e^b\,_\nu
-\,\,^0\Gamma^\sigma_{\mu\nu} e^a\,_\sigma \equiv 0\,.
\label{2-1-7}
\end{equation}
We lower the index $a$ in the equation above, multiply all terms by 
$e^{a\lambda}$, and obtain

\begin{equation}
e^{a\lambda} \partial_\mu e_{a\nu}=\,\,^0\Gamma^\lambda_{\mu\nu} -
e^{a\lambda}(\,^0\omega_{\mu ab}) e^b\,_\nu\,.
\label{2-1-8}
\end{equation}
The left hand side is the Weitzenb\"{o}ck connection, which we denote as
before by $\Gamma ^\lambda_{\mu\nu}$. Taking into account Eq. (\ref{2-1-2}),
we obtain the identity

\begin{equation}
\Gamma^\lambda_{\mu\nu}=\,\,^0\Gamma^\lambda_{\mu\nu} +
e^{a\lambda} K_{\mu ab}  e^b\,_\nu\,.
\label{2-1-9}
\end{equation}

With the help of the identities above we may write the scalar curvature 
density $eR(e)=e\,e^{a\mu}e^{b\nu}R_{ab\mu\nu}(^0\omega)$
constructed out of the Levi-Civita
connection $^0\omega_{\mu ab}$, in terms of the tetrad fields and the
torsion tensor $T_{a\mu\nu}=\partial_\mu e_{a\nu}-\partial_\nu e_{a\mu}$.
Taking into account Eq. (\ref{2-1-2}) in the expression of $eR(e)$, we obtain

\begin{equation}
eR(e)= -e({1\over 4}T^{abc}T_{abc}+{1\over 2}T^{abc}T_{bac}-T^aT_a)
+2\partial_\mu(eT^\mu)\,,
\label{2-1-10}
\end{equation}
where $T_\mu=T^\alpha\,_{\alpha\mu}$. 

Identity (\ref{2-1-10}) is also obtained by means of Eq. 
(\ref{2-1-9}). We first note that the curvature tensor constructed out
of the Weitzenb\"{o}ck connection vanishes identically,
$R_{\mu\nu\alpha\beta}(\Gamma)=0$. Then we consider the standard form of the 
scalar curvature density in terms of the metric tensor and the Christoffel 
symbols, $\sqrt{-g}g^{\mu\alpha}g^{\nu\beta}R_{\mu\nu\alpha\beta}(^0\Gamma)$,
make use of Eq. (\ref{2-1-9}), and eventually arrive at Eq. (\ref{2-1-10}).

We note finally that since the left hand side of Eq. (\ref{2-1-10}) is 
invariant under local SO(3,1) transformations, the right hand side of the 
equation (including the total divergence) is 
also invariant under the same transformations.

\subsection{The field equations of the TEGR and the gravitational 
energy-momentum tensor}

We introduce the tensor $\Sigma^{abc}$ defined by \cite{Maluf1}

\begin{equation}
\Sigma^{abc}={1\over 4}(T^{abc}+T^{bac}-T^{cab})+
{1\over 2}(\eta^{ac}T^b-\eta^{ab}T^c)\,,
\label{2-2-1}
\end{equation}
which yields the quadratic combination of the torsion tensor,

\begin{equation}
\Sigma^{abc}T_{abc}=
{1\over 4}T^{abc}T_{abc}+{1\over 2}T^{abc}T_{bac}-T^aT_a\,.
\label{2-2-2}
\end{equation}
Thus, identity (\ref{2-1-10}) may be rewritten as

\begin{equation}
eR(e)= -e\Sigma^{abc}T_{abc} +2\partial_\mu(eT^\mu)\,.
\label{2-2-3}
\end{equation}
Except for the total divergence, the quadratic scalar density 
$e\Sigma^{abc}T_{abc}$ is equivalent to the scalar curvature density
$eR(e)$. 

Therefore we define the Lagrangian density of the TEGR as \cite{Maluf1,Maluf7}

\begin{equation}
L(e)=-ke\Sigma^{abc}T_{abc} - {1\over c}L_M\,,
\label{2-2-4}
\end{equation}
where $L_M$ stands for the Lagrangian density of the matter fields, and 
$k=c^3/16 \pi G$ or, in natural units, $k=1/16 \pi$ ($G$ is the gravitational
constant). The absence in the Lagrangian density of the total divergence that
arises in the right hand side of Eq. (\ref{2-2-3}) prevents the invariance
of Eq. (\ref{2-2-4}) under local, arbitrary SO(3,1) transformations. 
However, if the matrices of the local SO(3,1) transformations fall off 
sufficiently fast at spacelike infinity, then the action integral
formed by the quadratic combination $e\Sigma^{abc}T_{abc}$ is invariant under
these special transformations \cite{Cho}. We assume in this review that the 
Lagrangian density above is constructed for asymptotically flat space-times. 
The Lagrangian density for more general space-times may be constructed with 
suitable surface terms (just like in the ordinary metrical formulation of
general relativity) that yield an invariant action integral $S$, whose 
variation $\delta S$ leads to the expected field equations. 

The field equations derived from arbitrary variations of $L(e)$
with respect to $e^{a\mu}$ are given by \cite{Maluf1,Maluf7}

\begin{equation}
e_{a\lambda}e_{b\mu}\partial_\nu (e\Sigma^{b\lambda \nu} )-
e (\Sigma^{b\nu}\,_aT_{b\nu\mu}-
{1\over 4}e_{a\mu}T_{bcd}\Sigma^{bcd} )={1\over {4k}}eT_{a\mu}\,,
\label{2-2-5}
\end{equation}
where $T_{a\mu}$ is defined by 
${{\delta L_M}/ {\delta e^{a\mu}}}=eT_{a\mu} $.
Although the Lagrangian density is not invariant under arbitrary SO(3,1)
transformations, the field equations (\ref{2-2-5}) are covariant under
local transformations of the SO(3,1) group.

The theory defined by the Lagrangian density (\ref{2-2-4}) is equivalent to
Einstein's general relativity because it can be shown that the left hand side
of Eq. (\ref{2-2-5}) is identically rewritten as
${1\over 2}e\left[ R_{a\mu}(e)-{1\over 2}e_{a\mu}R(e)\right]$. In order to 
prove the identity, it is easier to
start with the left hand side of Eq. (\ref{2-2-5}) and arrive at the latter 
expression. For this purpose, the following three identities are helpful. Let
us define $\,^0\omega^\mu\equiv \,\,^0\omega_\lambda\,^{\lambda\mu}$. The 
identities 

\begin{eqnarray}
T_\mu&=& -\,\,^0\omega_\mu\,, \nonumber \\
T_{\mu\lambda\nu}&=
& \,\,^0\omega_{\lambda\nu\mu}-\,\,^0\omega_{\nu\lambda\mu}\,, 
\nonumber \\
\Sigma_{\mu bc}&=&-{1\over 2} (\,\,^0\omega_{\mu bc}-e_{b\mu}\,\,^0\omega_c+
e_{c\mu}\,\,^0\omega_b)\,,
\label{2-2-7}
\end{eqnarray}
are useful in obtaining 
${1\over 2}e\left[ R_{a\mu}(e)-{1\over 2}e_{a\mu}R(e)\right]$ by means of 
algebraic manipulations of the left hand side of (\ref{2-2-5}). We note that 
by means of these identities, one may always transform the standard form of 
Einstein's equations for a general (non-asymptotically flat) space-time into
the field equations of the TEGR.

In Ref. \cite{Maluf11} it is shown that the coupling of a Dirac spinor field
with the gravitational field, in the framework of the Lagrangian density 
(\ref{2-2-4}), is consistent. The coupling is established by considering 
$\,^0\omega_{\mu ab}=-K_{\mu ab}$ in the covariant derivative of the Dirac
field. By using the resulting Dirac equation, it can be shown that the
energy-momentum tensor for the Dirac field is symmetric.

The indices in the field equation (\ref{2-2-5}) may be converted into 
space-time indices, and thus the left hand side of the latter equation 
becomes proportional to 
$(R_{\mu\nu}-{1\over 2}g_{\mu\nu}R)$. Consequently, a metric 
tensor $g_{\mu\nu}$ that is a solution of Einstein's equations is also a 
solution of Eq. (\ref{2-2-5}). For a given space-time metric, there exists an
infinity of allowed frames. Therefore, all physical results derived from
considerations of a space-time metric tensor, that is solution of Einstein's
equations, are valid in the present
formulation of the TEGR. In particular, the coupling of the gravitational 
field with the electromagnetic field may be established in the standard way
according to $eg^{\mu\alpha}g^{\nu\beta}F_{\mu\nu}F_{\alpha\beta}$, where
$F_{\mu\nu}$ is the Faraday tensor. Thus, the electromagnetic field may
couple to torsion, but in this context the concept of torsion is not the 
same as in the Einstein-Cartan theory (see the discussion in Section 7), where
torsion is normally considered as an additional geometrical quantity in a 
metric theory.

Equation (\ref{2-2-5}) may be rewritten as 

\begin{equation}
\partial_\nu(e\Sigma^{a\lambda\nu})={1\over {4k}}
e\, e^a\,_\mu( t^{\lambda \mu} + T^{\lambda \mu})\;,
\label{2-2-8}
\end{equation}
where $T^{\lambda\mu}=e_a\,^{\lambda}T^{a\mu}$ and
$t^{\lambda\mu}$ is defined by

\begin{equation}
t^{\lambda \mu}=k(4\Sigma^{bc\lambda}T_{bc}\,^\mu-
g^{\lambda \mu}\Sigma^{bcd}T_{bcd})\,.
\label{2-2-9}
\end{equation}
The tensor $\Sigma^{a\mu\nu}$ is antisymmetric in the last two indices,
$\Sigma^{a\mu\nu}=-\Sigma^{a\nu\mu}$, and from this property it  follows that
$\partial_\lambda \partial_\nu(e\Sigma^{a\lambda\nu})\equiv 0$. Therefore,

\begin{equation}
\partial_\lambda
\left[e\, e^a\,_\mu( t^{\lambda \mu} + T^{\lambda \mu})\right]=0\,.
\label{2-2-10}
\end{equation}
In the standard metrical formulation of general relativity, there is no 
equation that is equivalent to (\ref{2-2-10}). The equation above yields the 
continuity,  or balance equation,

\begin{equation}
{d\over {dt}} \int_V d^3x\,e\,e^a\,_\mu (t^{0\mu} +T^{0\mu})
=-\oint_S dS_j\,
\left[e\,e^a\,_\mu (t^{j\mu} +T^{j\mu})\right]\,,
\label{2-2-11}
\end{equation}
where the integration is carried out over a three-dimensional volume $V$,
bounded by the surface $S$.

The tensors $t^{\lambda\mu}$ and $T^{\lambda\mu}$ appear on the same footing
in Eqs. (\ref{2-2-10}) and (\ref{2-2-11}). We are led to interpret 
$t^{\lambda\mu}$ as the gravitational energy-momentum tensor, and the quantity
on the left hand side of Eq. (\ref{2-2-11}),

\begin{equation}
P^a=\int_V d^3x\,e\,e^a\,_\mu (t^{0\mu} 
+T^{0\mu})\,,
\label{2-2-12}
\end{equation}
as the total energy-momentum contained within the volume $V$ 
\cite{Maluf3,Maluf4}. In view of the field equation (\ref{2-2-8}), $P^a$ may 
be rewritten as 

\begin{equation}
P^a=-\int_V d^3x \partial_j \Pi^{aj}=-\oint_S dS_j\,\Pi^{aj}\,,
\label{2-2-13}
\end{equation}
where $\Pi^{aj}=-4ke\,\Sigma^{a0j}$. The expression above is the definition 
for the gravitational energy-momentum presented in Refs. \cite{Maluf3,Maluf4},
obtained in the framework of the vacuum field equations in Hamiltonian form. 
It is invariant under coordinate transformations of the three-dimensional 
space, under time reparametrizations and under global SO(3,1)
transformations. In vacuum, Eq. (\ref{2-2-13}) 
represents the gravitational energy-momentum vector $P^a=(E/c, {\bf P})$. We
will reconsider the gravitational energy-momentum vector in Section 5, after
the presentation of the Hamiltonian formulation. Expressions for the energy,
momentum and angular momentum of the gravitational field arise in the context
of the constraint equations of the Hamiltonian formulation of the theory, as
we will see in Section 5. The definition of the gravitational angular 
momentum, to be presented ahead, can only be obtained in the Hamiltonian 
framework.

We see that (\ref{2-2-10}) is a true energy-momentum conservation equation.
If we let $V\rightarrow \infty$, the right hand side of Eq. (\ref{2-2-11})
goes to zero if the relevant field quantities fall off sufficiently fast at
spacelike infinity. By inspecting the right hand side of Eq. (\ref{2-2-11}), 
we define \cite{Maluf10}

\begin{equation}
\Phi^a_g=\oint_S dS_j\,
\, (e\,e^a\,_\mu t^{j\mu})\,,
\label{2-2-14}
\end{equation}
as the gravitational energy-momentum flux, and

\begin{equation}
\Phi^a_m=\oint_S dS_j\,
\,( e\,e^a\,_\mu T^{j\mu})\,,
\label{2-2-15}
\end{equation}
as the energy-momentum flux of matter. Therefore the $a=(0)$ component of 
Eq. (\ref{2-2-11}) yields 

\begin{equation}
{{d P^{(0)}}\over {dt}}=-\Phi^{(0)}_g-\Phi^{(0)}_m\;.
\label{2-2-16}
\end{equation}

The expressions and definitions above are consequence of field equations
(\ref{2-2-5}) or (\ref{2-2-8}) only. No consideration is made to action 
integrals, surface terms or boundaries. 

The present formalism may be used to obtain the gravitational pressure on the
external event horizon of the Kerr black hole \cite{Maluf12}, for instance. 
In vacuum, the conservation equation (\ref{2-2-11}) is written as

\begin{equation}
{{dP^a}\over {dt}}=
-\oint_S dS_j\,
\left[e\,e^a\,_\mu t^{j\mu} \right]\,.
\label{2-2-17}
\end{equation}
Considering the field equation (\ref{2-2-8}), the right hand side of the 
equation above becomes

\begin{equation}
{{dP^a}\over {dt}}=
-4k\oint_S dS_j\,
\partial_\nu(e\Sigma^{a j\nu})\,.
\label{2-2-18}
\end{equation}
Restricting now the index $a$ to $a=(i)$, where $i=1,2,3$, we find

\begin{equation}
{{dP^{(i)}}\over {dt}}= -\oint_S dS_j\, \phi^{(i)j}
=-\oint_S dS_j \left[ e e^{(i)}\,_\mu t^{j\mu}\right] \,,
\label{2-2-19}
\end{equation}
where

\begin{equation}
\phi^{(i)j}=4k\partial_\nu(e\Sigma^{(i)j\nu}) \,.
\label{2-2-20}
\end{equation}
The left hand side of Eq. (\ref{2-2-19}) represents the momentum of the 
field divided by time, and therefore it has dimension of force ($\phi^{(i)j}$
should not be confused with $\phi_{ab}$ given by Eq. (\ref{1-10})). Since on 
the right hand side $dS_j$ is an element of area, we see that $-\phi^{(i)j}$ 
represents the pressure along the $(i)$ direction, over and element of area 
oriented along the $j$ direction. In Cartesian coordinates the index $j=1,2,3$
represents the directions $x,y,z$ respectively. In Ref. \cite{Maluf12} the
gravitational pressure on the external event horizon of the Kerr black hole
has been evaluated in the analysis of the thermodynamic relation
$TdS=dE + p\,dV$. 

\subsection{f($T$) theories of gravity}

The teleparallel framework allows the formulation of an interesting class of  
alternative theories of gravity, known as f($T$) theories, where 
\textquotedblleft{} f \textquotedblright {} is a 
functional of $T=\Sigma^{abc}T_{abc}$. One of the first attempts was the 
construction of a Born-Infeld type theory, with the purpose of arriving at 
regular and singularity-free solutions of the field equations, just like in 
the Born-Infeld formulation of electrodynamics. This approach was carried out
by Ferraro and Fiorini \cite{Ferraro}, who proposed the theory defined by
the Lagrangian density

\begin{equation}
L=-{{\lambda\,c^3}\over {16 \pi\,G}}e
\biggl[\sqrt{1+{{2\Sigma^{abc}T_{abc}}\over \lambda}}
-1 \biggr]\,,
\label{2-3-1}
\end{equation}
where $\lambda$ is a Born-Infeld parameter that controls the scale at which the
deformed solutions differ from the solutions of the standard theory (obtained
in the limit $\lambda \gg \Sigma^{abc}T_{abc}$). Ferraro and Fiorini 
investigated black hole solutions, and the spatially flat
Friedmann-Robertson-Walker cosmological model. 

Perhaps the most interesting application of f($T$) theories is the attempt 
to explain the accelerating expansion of the universe. Presently there is a 
variety of theoretical models that propose an explanation of the cosmic 
expansion, suggested by recent cosmological observations of Supernovas.
Modified teleparallel gravity allows an alternative understanding of this 
important problem (see, for instance, Refs. \cite{mod-tele}), without 
resorting to the dark energy concept, to inflationary models, to unimodular 
gravity or to gravity theories with a cosmological constant. One relevant 
feature of these models is that the field equations of the theory are always
second order differential equations, irrespective of the functional form of 
f($T$) (this feature is not shared by the corresponding f($R$) models, where 
$R$ is the scalar curvature). The Friedmann equations are slightly modified,
and may be solved numerically in order to yield very interesting results. The
field equations also allow the investigation of the existence of relativistic
stars in the framework of f($T$) theories \cite{Bohmer}, and of wormhole 
solutions in some viable models \cite{Jamil}.

\section{The Hamiltonian formulation of the TEGR}

The Hamiltonian formulation is of fundamental importance in the analysis of 
the structure of any physical theory. In field theory it reveals the  
existence of hyperbolic differential equations (time evolution equations),
of elliptic differential equations (constraint equations), of the dynamic and 
non-dynamic field quantities, and of the radiating degrees of freedom of
the theory. A well defined physical theory must necessarily have a well 
defined and consistent Hamiltonian formulation. 
The relevance of the Hamiltonian formulation of general relativity
is clear from the work of Arnowitt, Deser, Misner (ADM) \cite{ADM}. The ADM 
formulation is used in approaches to the quantization of the gravitational 
field, as well as in the establishment of the initial value 
problem for configurations like binary black holes, with the purpose
of investigating the time evolution of the system. With the use of numerical
analysis and computational tools, the Hamiltonian formulation allows the 
investigation of the strong-field, non-linear nature of the gravitational 
field.

The Hamiltonian formulation of the TEGR is formulated by means of the
following procedure. We start with the Lagrangian density (\ref{2-2-4}) and
make $L_M=0$. The idea is to write the Lagrangian density $L$ in the form
$p\dot{q} -H$, where $H$ is recognized as the Hamiltonian density. The
procedure requires the realization of the Legendre transform. As in the ADM
formulation, this is a non-trivial step. The procedure demands the ability
to identify the Lagrange multipliers as non-dynamic components of the
tetrad fields, and the primary constraints out of the components of the
momenta. The Hamiltonian formulation of the TEGR was first addressed in
Ref. \cite{Maluf1}, where Schwinger's time gauge condition was imposed on the
tetrad fields in order to simplify the calculations. The resulting Hamiltonian
formulation is very similar to the ADM formulation. In particular, the
constraints and constraint algebra resemble the corresponding expressions of
the ADM formulation.

The full Hamiltonian formulation of the TEGR was established in Ref. 
\cite{Maluf5}, but a refined formulation was presented in Ref. \cite{Maluf6}. 
The constraint algebra presented in the latter reference is similar to the 
algebra of the Poincar\'e group. The Hamiltonian formulation of unimodular 
gravity in the realm of the TEGR was investigated in Ref. \cite{Maluf6}. We
will dispense with the unimodular condition on the metric tensor, and follow 
Ref. \cite{Maluf6} in this short presentation of the Hamiltonian formulation 
of the TEGR.

In the present construction of the Hamiltonian formulation, we deal 
directly with the space-time components of both the tetrad fields and metric 
tensor. We do not carry out a 3+1 decomposition of the latter field 
quantities, i.e., the tetrad fields and the metric tensor are not projected on
three-dimensional spacelike hypersurfaces. From the Lagrangian density  
(\ref{2-2-4}) we obtain the momentum canonically conjugated to $e_{a\mu}$. It 
reads

\begin{equation}
\Pi^{a\mu}= {{\delta L}\over{\delta\dot{e}_{a\mu}}}=-4ke\,\Sigma^{a 0\mu}\;,
\label{3-1-1}
\end{equation}
where the dot over $e_{a\mu}$ represents the time derivative.
Given that $\Sigma^{abc} = -\Sigma^{acb}$, we have $\Pi^{a0} =0$, which
is a consequence of the fact that there is no time derivative of $e_{a0}$.
We refer the reader to Ref. \cite{Maluf6} for all details of this analysis.

We first obtain the primary Hamiltonian $H_0=\Pi^{ai} \dot{e}_{ai}-L$. It is
given by 

\begin{eqnarray}
H_{0}(e_{ai}, \Pi^{ai}, e_{a0}) &=&
-e_{a0}\partial_{i}\Pi^{ai} -
\frac{ke}{4g^{00}}\Big(g_{ik}g_{jl}P^{ij}P^{kl}-\frac{1}{2}P^{2}\Big)
\nonumber \\
&+& ke\Big(
\frac{1}{4}g^{ik}g^{jl}T^{a}\,_{ij}T_{akl} +
\frac{1}{2}g^{jl}T^{k}\,_{ij}T^{i}\,_{kl} \nonumber \\
&-& g^{il}T^{j}\,_{ji}T^{k}\,_{kl}\Big) \;. 
\label{3-1-13}
\end{eqnarray}
The quantity $P^{ij}$ is defined by 

\begin{equation}
P^{ki} = \frac{1}{ke}\Pi^{(ki)} - \Delta^{ki}\;,
\label{3-1-9}
\end{equation}
where 
$$\Delta^{ki} = -g^{0m}(g^{kj}T^{i}\,_{mj} + g^{ij}T^{k}\,_{mj} -
2g^{ik}T^{j}\,_{mj}) 
-(g^{km}g^{0i} + g^{im}g^{0k})T^{j}\,_{mj}\,.$$

The definition of the momenta $\Pi^{ai}$ leads to primary constraints
$\Gamma^{ab}=0$, 
\begin{equation}
\Gamma^{ab} = -\Gamma^{ba}=\Pi^{[ab]} + 4ke(\Sigma^{a0b} -
\Sigma^{b0a})\;,
\label{3-1-16}
\end{equation}
and to $\Pi^{a0}=0$. Secondary constraints
$C^a=0$ arise from the time evolution of the primary constraints 
$\Pi^{a0}$, i.e., by requiring that $\dot{\Pi}^{a0}$ vanishes weakly. The
constraints $\Gamma^{ab}$ do not yield secondary constraints.
The full Hamiltonian density is given by

\begin{equation}
H(e_{a\mu}, \Pi^{a\mu}, \lambda_{ab},\lambda_a)
= e_{a0}C^{a} + \lambda_{ab}\Gamma^{ab} + \lambda_{a}\Pi^{a0}\;,
\label{3-2-8}
\end{equation}
where $\lambda_{ab}$ and $\lambda_{a}$ are Lagrange multipliers that are
precisely determined by the evolution equations. The full expression of $C^a$
may be presented in a simplified form as \cite{Maluf6}

\begin{equation}
C^{a}= e^{a0}H_0  + e^{ai} H_i\;,
\label{3-2-4}
\end{equation}
where $H_i$ is defined by

\begin{equation}
H_i = -e_{ai}\partial_{k}\Pi^{ak} - \Pi^{ak}T_{aki}\;.
\label{3-2-5}
\end{equation}
It follows from Eq. (\ref{3-2-4}) that $e_{a0}C^{a} =  H_{0}$. It is 
important to observe, however, that the constraint $C^a$ may also be
rewritten as

\begin{equation}
C^a=-\partial_i \Pi^{ai} - h^a\,,
\label{3-2-44}
\end{equation}
where $h^a$ is obtained from Eqs. (\ref{3-1-13}) and (\ref{3-2-4}).
This form of $C^a$ will be crucial in the following section.

The Poisson brackets of the constraints $C^a$ and $\Gamma^{ab}$ is given
by \cite{Maluf6}

\begin{equation}
\{C^{a}(x), C^{b}(y)\} = 0\;, 
\label{3-3-7}
\end{equation}

\begin{equation}
\{C^{a}(x), \Gamma^{bc}(y)\} = \left(\eta^{ab}C^{c}
- \eta^{ac}C^{b}\right)\delta (x - y)\;, 
\label{3-3-8}
\end{equation}
\begin{equation}
\{\Gamma^{ab}(x), \Gamma^{cd}(y)\} = \left(
\eta^{ad}\Gamma^{bc} + \eta^{bc}\Gamma^{ad} 
- \eta^{ac}\Gamma^{bd} -
\eta^{bd}\Gamma^{ac}\right)\delta (x-y)\,.
\label{3-3-9}
\end{equation}
All Poisson brackets of the constraints $\Pi^{a0}$ with both $C^{a}$ and 
$\Gamma^{ab}$ vanish strongly.
In view of the constraint algebra above, we see that the constraints 
$C^a$, $\Gamma^{ab}$ and $\Pi^{a0}$ constitute a set of first class 
constraints. Thus, the Hamiltonian formulation of the TEGR is mathematically 
well established, and the initial value problem is well defined.

The constraints $C^a$ and $\Gamma^{ab}$ are labelled with SO(3,1) indices, 
and consequently the gravitational energy-momentum and angular momentum 
densities, to be discussed in the next Section, are also labelled with
these indices. Moreover, the algebra given by Eqs. (\ref{3-3-7}-\ref{3-3-9})
is very much similar to the algebra of the Poincar\'e group. These features 
justify the use of the SO(3,1) indices in labelling the constraints.

The physical degrees of freedom of the theory may be 
counted in the following way. The pair of dynamical field quantities 
$(e_{ai},\Pi^{ai})$ displays $12+12=24$ degrees of freedom. The $4+6$ first 
class constraints $(C^a, \Gamma^{ab})$ generate symmetries of the action, and 
thus they reduce 10+10=20 degrees of freedom. Therefore in the phase space of 
the theory there are 4 degrees of freedom, as expected. The action of the
constraints on the tetrad fields and on the metric tensor is 
explicitly discussed in Ref. \cite{Maluf6}.

\section{Energy, momentum and angular momentum of the gravitational field}

The definitions of the energy, momentum and angular momentum of the 
gravitational field constitute a long standing problem in the theory of 
general relativity. These definitions are necessary in order to have a
comprehensive understanding of the theory. The first approach to a solution of
this problem consisted in the derivation of 
energy-momentum pseudo-tensors. However, the 
solution presented by this approach is not satisfactory for at least two
important reasons. The first is that pseudo-tensors are not well defined with
respect to coordinate transformations. As a consequence, the results obtained 
via pseudo-tensors are ``valid'' only in one coordinate system. The second 
reason is that there are several pseudo-tensors available in the literature, 
and there is no explanation as to why one pseudo-tensor is better than another 
one. The principle of equivalence is sometimes invoked to justify the 
non-existence of a well defined expression for the gravitational energy 
density. The 
idea is that since one can transform an arbitrary metric tensor to the 
Minkowski metric tensor along any timelike worldline of an observer, a well 
defined expression for the gravitational energy density cannot exist, since
one may ``remove'' the gravitational field along this worldline. The problem
with this argument is that the transformation in consideration can be carried
out also along any spacelike trajectory, independently of whether the metric
tensor obeys any field equations. The reduction of the metric tensor to the 
Minkowski metric tensor along any worldline is a feature of differential
geometry, and is not a manifestation of any physical principle \cite{Babak}.
Moreover, this criticism does not apply to
the teleparallel framework, because the torsion tensor cannot be made to
vanish at a point in space-time by means of a coordinate transformation. 

An important step towards the concept of gravitational energy-momentum was 
provided by the work of Arnowitt, Deser and Misner \cite{ADM}. In this 
framework, the total gravitational
energy-momentum is given by surface integrals, constructed out of the 
components of the metric tensor at spatial infinity, and is valid only for
asymptotically flat space-times. The ADM energy-momentum first appeared in the 
construction of the Hamiltonian formulation of general relativity. It must be
present in the total Hamiltonian of the theory (the integral of the 
Hamiltonian and vector constraints, multiplied by the lapse and shift 
functions, respectively), so that the total Hamiltonian has well defined 
functional derivatives with respect to the phase space variables. In this 
case, the total
Hamiltonian generates the correct equations of motion \cite{Regge}. It is
important to mention that the search for the gravitational energy
density within the Hamiltonian formulation of general relativity was suggested
to follow from a canonical transformation of the phase space variables of the
ADM formulation, to new variables that would be classified as (i) embedding 
variables, and (ii) the true gravitational degrees of freedom \cite{Kuchar1}.
After this transformation, one would expect the constraints to be written as
$H_A=P_A+h_A=0$, $\,(A=0,1,2,3)$, 
where $P_A$ is the embedding momenta, and $h_A$ is the 
gravitational energy density and energy flux carried by the true gravitational
degrees os freedom \cite{Kuchar2}. However, this suggestion has never been 
implemented in the metrical ADM formulation.

It is clear from the analyses of the pseudo-tensors and of the total 
gravitational energy-momentum provided  by the ADM approach, that the 
gravitational energy density must be given by the second order derivatives of
the metric tensor. However, there does not exist a non-trivial, covariant 
expression constructed out of the metric tensor that yields, at the same time,
a scalar density that may be interpreted as the gravitational energy density, 
and the total ADM energy, when integrated over the whole three-dimensional
space. This is a limitation of the metrical formulation of general relativity.
It turns out that such expression exists in a theory formulated in
terms of the torsion tensor. 

In the TEGR, the field equations of the theory (Euler-Lagrange and first class
constraint equations) are interpreted as equations that define the energy, 
momentum and angular momentum of the gravitational field. We already verified
that, in the context of the Euler-Lagrange field equations, we may obtain 
definitions (\ref{2-2-12}) and (\ref{2-2-13}), together with the balance
equations (\ref{2-2-11}), (\ref{2-2-16}) and (\ref{2-2-18}), 
which establish the conservation
of the gravitational energy-momentum. In the Hamiltonian framework, a similar
feature takes place. The interpretation of a constraint equation as an energy
equation for a physical system is not a specific feature of the TEGR. It 
occurs, for instance, in the consideration of Jacobi's action 
\cite{Lanczos} for a parametrized non-relativistic particle. In order to 
make clear this feature, let us consider a particle of mass $m$ described in
the configuration space by generalized coordinates $q^i$, $i=1,2,3$. The 
particle is subject to the potential $V(q)$ and has constant energy $E$. 
Denoting $\dot{q}^i=dq^i/dt$, where $t$ is a monotonically increasing 
parameter between the (fixed) initial and end points of the path, the Jacobi
action integral for this particle can be written as \cite{BY1}

\begin{equation}
I=\int_{t_1}^{t_2} dt
\sqrt{m g_{ij}(q) \dot{q}^i \dot{q}^j}
\sqrt{2\lbrack E-V(q)\rbrack}\,.
\label{4-1-1}
\end{equation}

\noindent The action is extremized by varying the configuration space path
and requiring $\delta q(t_1)=\delta q(t_2)=0$. We may simplify the integrand
by writing $dt \sqrt{m g_{ij}\dot{q}^i \dot{q}^j}=
\sqrt{m g_{ij} dq^i dq^j}$, which shows that the action is invariant
under reparametrizations of the time parameter $t$. Thus, in Jacobi's 
formulation of the action principle, it is the energy $E$ of the particle that
is fixed, not its initial and final instants of time. In view of the time 
reparametrization of the action integral, the Hamiltonian constructed out of 
the Lagrangian above vanishes identically, which is a feature of 
reparametrization invariant theories. If we denote $p_i$ as the momenta 
conjugated to $q^i$, we find 
$p_i=(g_{ij} \dot{q}^j/\sqrt{m}\,)\sqrt{2(E-V)/\dot{q}^2}$ 
(where $\dot{q}^2=g_{kl}\dot{q}^k\dot{q}^l$), which leads to the constraint

\begin{equation}
C(q,p)\equiv {1\over {2m}}g^{ij}p_i p_j + V(q)-E\approx 0
\label{4-1-2}
\end{equation}

\noindent The equation of motion obtained from the action integral has to be
supplemented by the constraint equation $C=0$, in order to be equivalent with
Newton's equation of motion with fixed energy $E$ \cite{BY1}. Therefore, we 
see that the constraint equation defines the energy of the particle. This is
exactly the feature that takes place in the TEGR: the definitions of the 
energy-momentum and angular momentum of the gravitational field arise from 
the constraint equations of the theory \cite{Maluf4,Maluf13}. These definitions
are viable as long as they yield consistent values in the consideration of 
relevant and well understood gravitational field configurations.

\subsection{Gravitational energy-momentum}

Let us consider the expression of the constraint equation $C^a=0$, where $C^a$
is given by Eq. (\ref{3-2-44}). The first term on the right hand side of Eq.
(\ref{3-2-44}) is $-\partial_i \Pi^{ai}$. We recall that the momenta $\Pi^{ai}$
is a density and reads $\Pi^{ai}=-4ke\,\Sigma^{a0i}$, according to Eq.
(\ref{3-1-1}). In the metrical formulation of general relativity there does 
not exist any quantity of the type of $-\partial_i \Pi^{ai}$, i.e., a 
non-trivial total divergence. The emergence of a density in the form of
a total divergence is the  motivation to consider the integral form of the 
constraint equation $C^a=0$,

\begin{equation}
-\partial_i \Pi^{ai} = h^a\,,
\label{4-2-2}
\end{equation}
as an equation for the gravitational energy-momentum. This is exactly the 
argument presented in Refs. \cite{Maluf2,Maluf4}. Therefore we define

\begin{equation}
P^a= - \int_V d^3x \partial_i \Pi^{ai} =-\oint_S dS_i\,\Pi^{ai}\,,
\label{4-2-3}
\end{equation}
as the total gravitational energy-momentum. This is the precisely expression 
(\ref{2-2-13}), obtained in the realm of the Lagrangian formulation. We recall
that Eq. (\ref{4-2-3}) was first presented in Ref. \cite{Maluf3}. The components
of the vector $P^a$ are $({E/c}, {\bf P})$.
If we assume that the tetrad fields satisfy asymptotic boundary conditions,

\begin{equation}
e_{a\mu} \simeq \eta_{a\mu}
+ {1\over 2}h_{a\mu}(1/r)\,,
\label{4-2-4}
\end{equation}
at spatial infinity, i.e., in the limit $r \rightarrow \infty$, then the total
gravitational energy $E=cP^{(0)}$ is the ADM energy \cite{Maluf4},

\begin{equation}
E ={{c^4}\over {16\pi G}}\int_{S\rightarrow \infty}dS_k(\partial_i
h_{ik}-\partial_k h_{ii}) = E_{ADM}\,.
\label{4-2-5}
\end{equation}

Definition (\ref{4-2-3}) has been applied to several configurations of the 
gravitational field, and all results are consistent. In this review we will
reconsider only one major result that follows from Eq. (\ref{4-2-3}). In the
next Section we will review the application of Eq. (\ref{4-2-3}) to the Kerr
space-time. It is important, however,
to address the problem of regularization of definition (\ref{4-2-3}).

Most sets of tetrad fields that are adapted to ordinary observers satisfy the
asymptotic boundary conditions (\ref{4-2-4}). It is clear that when we enforce
the vanishing of the physical parameters of the metric tensor, such as mass,
angular momentum and charge, the space-time in consideration is reduced to the
flat space-time, and in this case one expects that the torsion tensor
components $T_{a\mu\nu}$ vanish. Indeed, all components $T_{a\mu\nu}$ vanish if 
they are obtained from tetrad fields that satisfy (\ref{4-2-4}), when we
require the vanishing of the physical parameters.

However, the tetrad fields do not always have the same asymptotic 
behaviour of the metric tensor. When this is the case, we may have
$T_{a\mu\nu}\ne 0$ even for the flat space-time. One example is given by the
following set of tetrad fields,

\begin{equation}
E^a\,_\mu(t,r,\theta,\phi)=\pmatrix{1&0&0&0\cr
0&1&0&0\cr
0&0&r&0\cr
0&0&0&r\sin\theta\cr}\,.
\label{4-2-6}
\end{equation}
The tetrad fields above yield the line element for the flat space-time in
spherical coordinates. From this set of tetrad fields we obtain three
non-vanishing components: $T_{(2)12}=1$, $T_{(3)13}=\sin\theta$, and 
$T_{(3)23}=r\cos\theta$. By transforming Eq. (\ref{4-2-6}) into Cartesian
coordinates, we clearly see that the latter does not display the 
boundary conditions given Eq. (\ref{4-2-4}) \cite{Maluf14}.

We will denote the set of flat tetrads that displays the feature above as
$E^a\,_\mu$, and the momenta constructed out of $E^a\,_\mu$ by $\Pi^{ai}(E)$.
The regularized form of the gravitational energy-momentum $P^a$ is defined by
\cite{Maluf14}

\begin{equation}
P^a=-\int_V d^3x\,\partial_k\lbrack\Pi^{ak}(e) - \Pi^{ak}(E)\rbrack\;.
\label{4-2-8}
\end{equation}
This definition guarantees that the energy-momentum of the flat space-time 
always vanishes. The tetrad fields $E^a\,_\mu$ are obtained from the physical
fields $e^a\,_\mu$ by just requiring the vanishing of the parameters 
($m,a,q, \dots$). We remark that regularized expressions like Eq. 
(\ref{4-2-8}) are useful in the investigation of cosmological models, where one
does not dispose of asymptotic boundary conditions \cite{Maluf14-1}.

The evaluation of definition (\ref{4-2-3}) is carried out in the configuration
space. The definition is invariant under (i) general coordinate 
transformations of the three-dimensional space, (ii) time reparametrizations,
and (iii) covariant under global SO(3,1) transformations. The non-covariance 
of Eq. (\ref{4-2-3}) under the local SO(3,1) group reflects the frame
dependence of the definition. In the TEGR each set of tetrad fields is 
interpreted as a reference frame in space-time. Integral quantities like 
$P^a$ cannot be covariant under local SO(3,1) transformations.

Invariance of the field quantities under local SO(3,1) 
(Lorentz) transformations imply that the measurement of these quantities is 
the same in inertial and accelerated  frames. This is not an expected feature
of concepts such as energy, momentum and angular momentum. The energy is 
always the zero component of an energy-momentum vector. It cannot be invariant
under any type of SO(3,1) transformation.

It is worthwhile to recall a simple physical situation in which the frame 
dependence of the gravitational energy-momentum takes place. For this purpose 
we consider a black hole of mass $m$ and an observer that is very distant from
the black hole. The black hole will appear to this observer as a 
particle of mass $m$, with energy $E=cP^{(0)}=mc^2$. 
The parameter $m$ is the rest mass of the black 
hole, i.e., the mass of the black hole in the frame where the black hole is at
rest. If, however, the black hole is moving at velocity $v$ with respect to 
the observer, then its total gravitational energy will be $E=\gamma m c^2$, 
where $\gamma=(1-v^2/c^2)^{-1/2}$. The gravitational energy is indeed the 
zero component of the gravitational energy-momentum vector.
This example is a consequence of the special theory of relativity, and 
demonstrates the frame dependence of the gravitational energy-momentum. 
The frame dependence is not restricted to observers at spacelike 
infinity. It holds for observers located everywhere in the three-dimensional
space.

Finally we mention that the evaluation of $P^a$ in a freely falling frame in
the Schwarzschild space-time leads to a vanishing gravitational 
energy-momentum, i.e., $P^a=0$ \cite{Maluf8}. This result is in agreement 
with the standard description of the principle of equivalence, since
the local effects of gravity are not measured by an observer in free fall. 
Such observer cannot measure its own gravitational acceleration. The tetrad
fields that establish the frame of an observer in free fall  
is related to stationary frames, for instance, by a frame transformation, 
not by a coordinate transformation.

\subsection{Gravitational angular momentum}

In the TEGR the definition of the gravitational angular momentum is also
obtained from the constraint equations of the theory, in similarity to the
definition of the gravitational energy-momentum discussed in Section 5.1.
The primary constraints $\Gamma^{ab}$ in the Hamiltonian density 
yield the equations $\Gamma^{ab}=0$, or

\begin{equation}
2\Pi^{[ab]} + 4ke(\Sigma^{a0b} - \Sigma^{b0a})=0\;.
\label{4-3-1}
\end{equation}
Therefore we define the gravitational angular momentum density as

\begin{equation}
M^{ab}=2\Pi^{[ab]}=-4ke(\Sigma^{a0b}-\Sigma^{b0a})\,,
\label{4-3-2}
\end{equation}
and the total angular momentum of the gravitational field, contained 
within a volume $V$ of the three-dimensional space, according to
\cite{Maluf13,Maluf15}

\begin{equation}
L^{ab}=-\int_V d^3x\; M^{ab}\,.
\label{4-3-3}
\end{equation}
The expression above may be calculated from the field quantities in
the configuration space of the theory. In contrast to the expression of the 
gravitational energy-momentum, Eq. (\ref{4-3-3}) does not arise in the form
of a total divergence. 

In the Newtonian description of classical mechanics, the angular momentum of
the source is frame dependent. This feature also holds in relativistic 
mechanics. If the angular momentum of the source in general is frame 
dependent, it is reasonable to consider that the 
angular momentum of the field is frame dependent as well. Differently from 
other definitions of gravitational angular momentum, that are formulated in 
terms of surface integrals at spacelike infinity and depend only on the 
asymptotic behaviour of the metric tensor, the definition considered here 
naturally depends on the frame, since it is covariant under global 
SO(3,1) transformations of the tetrad fields. In the present framework,
observers that are in rotational motion around the rotating source measure 
the angular momentum of the gravitational field differently from static 
observers. Rotating and static observers also obtain different values for the
angular momentum of the source. 
In Newtonian mechanics, the angular momentum of the source,
in the frame of observers that co-rotate with the source, vanishes.

Let us consider a general line element for a space-time with axial symmetry,

\begin{equation}
ds^2=g_{00}dt^2+g_{11}dr^2+g_{22}d\theta^2+g_{33}d\phi^2
+2g_{03}d\phi\, dt\,,
\label{4-3-5}
\end{equation}
where all metric components depend on the spherical coordinates $r$ and 
$\theta$: $g_{\mu\nu}=g_{\mu\nu}(r,\theta)$. Here we will adopt $c=1$.
One relevant frame is determined by the set of tetrad fields adapted to 
stationary observers. This frame is established by the conditions

\begin{equation}
e_{(0)}\,^i=u^i=0\,,
\label{4-3-6}
\end{equation}
in spherical coordinates. We also choose the 
$e_{(3)}\,^\mu$ component to be oriented asymptotically 
($r\rightarrow \infty$) with the unit vector 
$\hat{\bf z}$ along the $z$ axis, namely,

\begin{equation}
e_{(3)}\,^\mu (t,x,y,z)
\cong (0,0,0,1)\,.
\label{4-3-9}
\end{equation}
The resulting frame in $(t,r,\theta,\phi)$ coordinates reads \cite{Maluf15}

\begin{equation}
e_{a\mu}=\pmatrix{-A&0&0&-C\cr
0&\sqrt{g_{11}} \,\sin\theta \cos\phi&
\sqrt{g_{22}} \cos\theta \cos\phi & -D\, r \sin\theta \sin\phi\cr
0& \sqrt{g_{11}}\, \sin\theta \sin\phi&
\sqrt{g_{22}} \cos\theta \sin\phi &  D\, r \sin\theta \cos\phi\cr
0& \sqrt{g_{11}}\, \cos\theta & -\sqrt{g_{22}}\sin\theta&0}\,.
\label{4-3-10}
\end{equation}
The functions $A, C$ and $D$ are defined by 

\begin{eqnarray}
A(r,\theta)&=& (-g_{00})^{1/2}\,, \ \ \ \ \ \
C(r,\theta)=-{{g_{03}}\over{(-g_{00})^{1/2}}}\,, \nonumber \\
D(r,\theta)
&=&\biggl[ {{-\delta} \over {(r^2\sin^2\theta) g_{00}}}
\biggr]^{1/2} \,,
\label{4-3-11}
\end{eqnarray}
where $\delta=g_{03}g_{03}-g_{00}g_{33}$.

We use definition (\ref{4-3-2}) to calculate the components of $M^{ ab}$. It
is possible to verify that only two components are non-vanishing. These 
components eventually arise as total divergences. We find \cite{Maluf15}

\begin{equation}
M^{(1)(2)}= 2k\biggl[
\partial_1\biggl(
{{g_{03}\sqrt{g_{22}}\,\sin\theta}\over \sqrt{-g_{00}}}\biggr)+
\partial_2 \biggl(
{{g_{03}\sqrt{g_{11}}\,\cos\theta}\over \sqrt{-g_{00}}}\biggr)
\biggr]\,,
\label{4-3-12}
\end{equation}
and

\begin{eqnarray}
M^{(0)(3)}&=& 2k\biggl[
\partial_1
\biggl( {{\delta^{1/2}\sqrt{g_{22}}\,\cos\theta}
\over {\sqrt{-g_{00}} }} \biggr) -\partial_2
\biggl( {{ \delta^{1/2}\sqrt{g_{11}}\,\sin\theta }\over 
{ \sqrt{-g_{00}} }} \biggr) \biggr]\,.
\label{4-3-13}
\end{eqnarray}
In order to obtain the total angular momentum of the gravitational field,
in the frame determined by Eq. (\ref{4-3-10}), we evaluate the integral of
Eq. (\ref{4-3-12}) as a surface integral, such that the surface of integration 
$S$, determined by the conditions $r=$ constant, is located at spacelike
infinity. We obtain

\begin{eqnarray}
L^{(1)(2)} =-\int d^3x\,M^{(1)(2)}= -2k
\oint_{S\rightarrow \infty}\,d\theta d\phi \biggl(
{{g_{03}\sqrt{g_{22}}\,\sin\theta}\over \sqrt{-g_{00}}}\biggr)\,.
\label{4-3-14}
\end{eqnarray}
We may then verify whether, for a given space-time metric tensor, the total
gravitational angular momentum is finite, vanishes or diverges. If 
$g_{\mu\nu}$ is given in spherical coordinates and if the following asymptotic
behaviour is verified,

\begin{eqnarray}
g_{03}& \cong & O(1/r)+\cdots \nonumber \\
g_{22}& \cong & r^2 + O(r)+ \cdots \nonumber \\
-g_{00}& \cong & 1 + O(1/r) + \cdots\,,
\label{4-3-15}
\end{eqnarray}
then expression (\ref{4-3-14}) will be finite. 

In Ref. \cite{Maluf15} a specific model for a rotating neutron star was 
investigated in detail. It was found that the angular momentum of the field is
given in terms of the angular momentum of the source $J_S$ according to the 
equation $L^{(1)(2)}=(2/3) J_S$. This result seems to be general, and is also
verified for the Kerr space-time, in the frame of static observers,
where we find $L^{(1)(2)}=(2/3) ma=(2/3) J$.

The quantity $L^{(0)(3)}$ is interpreted as the gravitational center of mass 
moment. It vanished for the rotating neutron star investigated in Ref. 
\cite{Maluf15}. The model determined by (\ref{4-3-5}) is arbitrary in the 
sense that the metric tensor depends arbitrarily 
on $\theta$. In view of the axial symmetry of the model, it is natural that 
the gravitational center of mass vanishes along the $x$ and $y$ directions,
but because of the $\theta$ dependence of the 
metric tensor, the integral of (\ref{4-3-13}) does not vanish in general.

In similarity to the definition of the regularized gravitational 
energy-momentum,
we may also establish a definition for the regularized gravitational angular
momentum. In view of Eq. (\ref{4-2-8}), we may extend the definition of the
gravitational angular momentum as

\begin{equation}
L^{ab}=-\int_V d^3x\; \lbrack M^{ab}(e) -M^{ab}(E) \rbrack\,,
\label{4-3-15-1}
\end{equation}
where $E^a\,_\mu$ is defined exactly as in Section 5.1.

With the purpose of analysing the frame dependence of the gravitational 
angular momentum, let us consider the general (and simple) form of the line
element that describes a rotating neutron star, for instance. This line 
element is given by \cite{Maluf15}

\begin{equation}
ds^2=-\alpha^2dt^2+\beta^2dr^2+r^2d\theta^2+r^2\sin^2\theta
(d\phi-\Omega dt)^2\,,
\label{4-3-16}
\end{equation}
where $\alpha$, $\beta$ and $\Omega$ are functions of the radial coordinate 
$r$ only. Again we adopt $c=1$.
The radius of the star is denoted by $R$. These quantities are 
defined for the interior ($r<R$) and for the exterior ($r>R$) of the star.
In the exterior region we have $\Omega(r)=2GJ_S/ r^3$ . 
Instead of adopting the frame determined by Eq. (\ref{4-3-6}), let us consider
a frame that satisfies Schwinger's time gauge condition,

\begin{equation}
e_{(i)}\,^0=0\,,
\label{4-3-18}
\end{equation}
together with condition (\ref{4-3-9}). In $(t,r,\theta,\phi)$ coordinates, the
frame reads \cite{Maluf15}

\begin{equation}
e_{a\mu}=\pmatrix{-\alpha&0&0&0\cr
\Omega r\,\sin\theta\,\sin\phi&\beta\,\sin\theta\,\cos\phi&
r\,\cos\theta\,\cos\phi&-r\,\sin\theta\,\sin\phi\cr
-\Omega r\,\sin\theta\,\cos\phi&\beta\,\sin\theta\,\sin\phi&
r\,\cos\theta\,\sin\phi&r\,\sin\theta\,\cos\phi\cr
0&\beta\,\cos\theta&-r\,\sin\theta&0}\,,
\label{4-3-19}
\end{equation}
This frame is adapted to the field of observers whose velocity 
$e_{(0)}\,^\mu$ is given by

\begin{equation}
e_{(0)}\,^\mu(t,r,\theta,\phi)=
{1\over \alpha}(1,0,0,\Omega(r)\,)\,.
\label{4-3-20}
\end{equation}
$\Omega(r)$ is the dragging velocity of inertial frames that rotate under the
action of the neutron star. The expression above for $e_{(0)}\,^\mu$ describes
the velocity field of observers that are dragged 
in circular motion around the star. It turns 
out that the angular momentum of the gravitational field calculated out of 
(\ref{4-3-19}) vanishes \cite{Maluf15}, 

\begin{equation}
L^{(1)(2)}=0\,,
\label{4-3-21}
\end{equation}
since $M^{(1)(2)}=0$. In fact, all
components of $M^{ab}$ vanish. This class of observers is known in the
literature as zero angular momentum observers (ZAMOs). They follow 
trajectories with constant radial coordinate $r$ and with angular velocity 
given by the dragging velocity of inertial frames.

The result given by Eq. (\ref{4-3-21}) shows that observers that are in 
rotational motion around the rotating source measure the gravitational angular
momentum differently from static observers. An explanation for this result 
must take into account the angular momentum of the source, which is different 
for observers at rest and for those that rotate around the source. In the 
Newtonian theory, the angular momentum of the source, in the frame
of observers that rotate at the same angular frequency, vanishes. We know that
this feature holds for a rigid body in Newtonian mechanics, where the angular 
momentum depends not only on the frame, but also on the origin of the frame.
Observers whose 
angular velocity around the rotating source is the same as the dragging 
velocity $\Omega(r)$ do not measure this dragging velocity (and possibly 
other dragging effects), and therefore for these observers the gravitational
angular momentum vanishes. 

\subsection{The algebra of the Poincar\'{e} group}

A very interesting property of definitions (\ref{4-2-3}) and (\ref{4-3-3})
for the gravitational energy-momentum $P^a$ and angular momentum $L^{ab}$,
respectively, is that these quantities 
satisfy the algebra of the Poincar\'e group in the phase space of the theory.
The functional derivatives of $P^a$ and $L^{ab}$ are 

$${{\delta L^{ab}}\over {\delta e_{ck}(z)}}=
\eta^{ac} \Pi^{bk}(z) - \eta^{bc} \Pi^{ak}(z)\,, \ \ \ \
{{\delta L^{ab}}\over {\delta \Pi_{ck}(z)}}=
-\delta^a_c e^b\,_k(z) + \delta^b_c e^a\,_k(z)\,,$$

$${{\delta P^a} \over {\delta e_{ck}(z)}}=0\,, \ \ \ \ 
{{\delta P^a} \over {\delta \Pi_{ck}(z)}}=-
\int d^3x \delta^a_c {\partial \over {\partial x^k}}\delta^3(x-z)\,. $$
In view of the relations above it is not difficult to arrive at the following
Poisson brackets,

\begin{eqnarray}
\lbrace P^a , P^b \rbrace &=& 0\,, \nonumber \\
\lbrace P^a , L^{bc} \rbrace &=& \eta^{ab} P^c-\eta^{ac} P^b 
\,, \nonumber \\
\lbrace L^{ab}, L^{cd} \rbrace &=&
\eta^{ac}L^{bd} +\eta^{bd}L^{ac} -\eta^{ad}L^{bc}-\eta^{bc}L^{ad}
\,.
\label{4-4-2}
\end{eqnarray}
It is known that quantities that satisfy the Poincar\'e algebra are intimately
related to energy-momentum and angular momentum. Therefore the Poincar\'e
algebra for $P^a$ and $L^{ab}$ confirms the consistency of the 
definitions.

\section{The Kerr space-time}

The definitions of the energy, momentum and angular momentum obtained in the 
TEGR may be applied to any configuration of the gravitational field, and also 
to cosmological models. In this review we discuss one relevant application, 
and consider the Kerr space-time. 
We will evaluate the energy contained within the external event horizon of the
Kerr black hole. This is a configuration where the concept of localization of
gravitational energy is very clear. No form of energy can escape from the 
external event horizon of the black hole, not even in the Penrose process of 
extraction of energy from black holes, for instance. This energy is related
to the irreducible mass of the black hole. This subject has already 
been investigated in the context of several definitions of gravitational 
energy. We will assume $c=1$. In spherical 
coordinates, the Kerr space-time is established by the line element

\begin{eqnarray}
ds^2&=&
-{{\psi^2}\over {\rho^2}}dt^2-{{2\chi\sin^2\theta}\over{\rho^2}}
\,d\phi\,dt
+{{\rho^2}\over {\Delta}}dr^2 \nonumber \\
&{}&+\rho^2d\theta^2+ {{\Sigma^2\sin^2\theta}\over{\rho^2}}d\phi^2\,,
\label{5-1-1}
\end{eqnarray}
with the following definitions:

\begin{eqnarray}
\Delta&=& r^2+a^2-2mr\,, \nonumber \\
\rho^2&=& r^2+a^2\cos^2\theta \,,  \nonumber \\
\Sigma^2&=&(r^2+a^2)^2-\Delta a^2\sin^2\theta\,, \nonumber \\
\psi^2&=&\Delta - a^2 \sin^2\theta\,, \nonumber \\
\chi &=&2amr\,.
\label{5-1-2}
\end{eqnarray}

We consider initially a stationary Kerr black hole with mass $m$ and angular 
momentum per unit mass $a=J/m$. In the Penrose process \cite{Penrose}
of extraction of energy of rotating black holes, the initial mass $m$ and 
angular momentum $J$ of the black hole vary by $dm$ and $dJ$, respectively, 
such that $dm -\Omega_H dJ \ge 0$, where $\Omega_H$ is the angular velocity 
of the external event horizon of the black hole,
$\Omega_H=a /(2mr_+) =a /( a^2+r_+^2)$, and 
$r_+$ is the radius of the external event horizon:
$r_+=m+\sqrt{m^2-a^2}$. In the Penrose process, the variation of the area $A$
of the black hole satisfies $dA \ge 0$. In the final stage of an idealized 
process, the mass of the black hole becomes the irreducible mass $m_{irr}$ 
\cite{CR}, defined by the relation $m^2=m_{irr}^2+J^2/(4m_{irr}^2)$, and
the Kerr black hole becomes a Schwarzschild black hole. The irreducible mass
is given by $m_{irr}={1\over 2} \sqrt{r_+^2+ a^2}$.
An analysis of various gravitational energy expressions for the Schwarzschild
and Kerr black holes has been carried out in Ref. \cite{Bergqvist}. 
Considering all known expressions for gravitational energy, it was concluded
that the energy contained within the event horizon of the Schwarzschild black 
hole is $2m$. One would expect that in the final stage of the Penrose process,
the energy contained within the external event horizon of the Kerr black hole
(which becomes a non-rotating black hole) would be $2m_{irr}$. However, none
of the expressions analysed in Ref. \cite{Bergqvist} yield $2m_{irr}$. 
The present definition for the gravitational energy yields an expression for
the energy contained within the external event horizon of the Kerr black hole
that is strikingly close to $2m_{irr}$. Let us first establish the frame.

The frame must be defined such that the radial coordinate $r$ runs from $r_+$
to infinity, i.e., the frame must be defined in the whole region outside the 
external event horizon, and consequently inside the ergosphere of the black 
hole. The ergosphere is defined by the region between the external event 
horizon, characterized by $r_+\,$, and $r_+^*=m+ \sqrt{m^2-a^2\cos^2 \theta}$.
The values of $r_+^*$ determine the 
external boundary of the ergosphere. We know that it is not
possible to establish a static frame inside the ergosphere, because in this
region all observers are necessarily dragged in circular motion by the 
gravitational field. The four-velocity of observers that circulate around the
black hole, outside the external horizon, under the action of the 
gravitational field of the Kerr space-time, is given by 

\begin{equation}
u^\mu(t,r,\theta,\phi)
={{\rho \Sigma}\over{(\psi^2\Sigma^2+\chi^2\sin^2\theta)^{1/2}}}
(1,0,0,{\chi \over {\Sigma^2}})\,,
\label{5-1-5}
\end{equation}
where all functions are defined in Eq. (\ref{5-1-2}). It is possible to show 
that if we restrict the radial coordinate to $r=r_+$, the $\mu=3$ component of
Eq. (\ref{5-1-5}) becomes
$ \chi/ \Sigma^2= a /2mr_+ = \Omega_H$. The quantity
$\omega(r)= -g_{03} /g_{33} = \chi /\Sigma^2$ 
is the dragging velocity of inertial frames.

The tetrad fields (i) that are adapted to observers whose four-velocities are
given by Eq. (\ref{5-1-5}), i.e., for which $e_{(0)}\,^\mu =u^\mu$, 
and consequently defined in the region $r\ge r_+$, (ii) whose $e_{(i)}\,^\mu$
components in Cartesian coordinates are asymptotically 
oriented along the unit vectors $\hat{\bf x}$, $\hat{\bf y}$, 
$\hat{\bf z}$, and (iii) that is asymptotically flat, is given by

\begin{equation}
e_{a\mu}=\pmatrix{-A&0&0&0\cr
B\sin\theta\sin\phi
&C\sin\theta\cos\phi& D\cos\theta\cos\phi&-E\sin\theta\sin\phi\cr
-B\sin\theta\cos\phi
&C\sin\theta\sin\phi& D\cos\theta\sin\phi& E\sin\theta\cos\phi\cr
0&C\cos\theta&-D\sin\theta&0}\,,
\label{5-1-7}
\end{equation}
where

$$A= {{(g_{03}g_{03}-g_{00}g_{33})^{1/2}}\over{(g_{33})^{1/2}}}\,, \ \ \ \ 
B=-{{ g_{03}}\over {(g_{33})^{1/2} \sin\theta}}\,,$$

$$C=(g_{11})^{1/2}\,, \ \ \ \  
D=(g_{22})^{1/2}\,, \ \ \ \ 
E= {{(g_{33})^{1/2}}\over {\sin\theta}}\,.$$
These tetrad fields are the unique configuration that satisfies the above
conditions, since six conditions are imposed on $e^a\,_\mu$. It satisfies
Schwinger's time gauge condition $e_{(i)}\,^0=0$. Therefore we 
may evaluate the gravitational energy contained within any surface $S$ 
determined by the condition $r\ge r_+$, and in particular for $r=r_+$.
Expression (\ref{5-1-7}) is precisely the same set of 
tetrad fields (Eq. (4.9)) considered in Ref. \cite{Maluf4}. This frame allows
observers to reach the vicinity of the external event horizon of the Kerr 
black hole.

The energy contained within the external event horizon of the black hole is
calculated by means of the $a=(0)$ component of Eq. (\ref{4-2-3}),

\begin{equation}
P^{(0)}=E=-\oint_S dS_i \Pi^{(0)i}=
-\oint_S d\theta d\phi\,\Pi^{(0)1}(r,\theta,\phi)\,.
\label{5-1-9}
\end{equation}
$S$ is a surface of constant radius determined by the condition $r=r_+$
After a number of algebraic calculations we obtain \cite{Maluf4}

\begin{equation}
E=m\biggl[ {\sqrt{2p}\over 4}+{{6p-k^2}\over {4k}}
\ln \biggl({{\sqrt{2p}+k}\over p}\biggl) \biggr]\,.
\label{5-1-10}
\end{equation}
The quantities $p$ and $k$ are defined by

$$p=1+\sqrt{1-k^2}\,, \ \ \ \ \ \ a= k\, m\,, \ \ \ \ \ \ 
0\le k \le 1\,.$$
The dimensionless parameter $k$ above should not be confused with 
$k=c^3/16\pi G$ in Eq. (\ref{2-2-4}). 
Equation (\ref{5-1-10}) is functionally different from  
$2m_{irr}=\sqrt{r_+^2+ a^2}$. However, the two expressions are very similar,
as we can verify in Fig. 1.

\begin{figure}[h]
\centering
\includegraphics[width=0.6\textwidth]{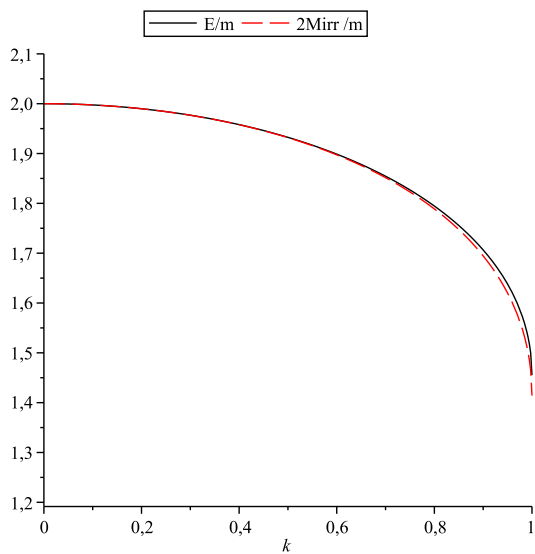}
\caption{Gravitational energy contained within the external event horizon of
the Kerr black hole, as function of the angular momentum per unit mass.
The upper figure represents $E/m$, and the lower one $2m_{irr}/m$. 
For a static or rotating black hole, the maximum value of E/m is 2.}
\end{figure}

In Fig. 1 we have plotted (i) $E/m$, where $E$ is given by Eq. 
(\ref{5-1-10}), and (ii) $2m_{irr}/m$, as functions of $a/m=k$. The 
curves are parametrized by $k$, which varies from 0 to 1. The upper curve
represents Eq. (\ref{5-1-10}), and the lower curve represents $2m_{irr}$. The
almost coincidence between the two expressions is striking, and is one major
achievement of definition (\ref{4-2-3}). It shows that Eq. (\ref{5-1-10}) is
in very close agreement with $2m_{irr}$, as expected. The result also supports
the idea of localization of the gravitational energy.

We conclude that the definition of gravitational energy is physically 
acceptable. The definition must be evaluated in the frame adapted to 
distinguished observers in space-time.

\section{Final remarks}

In this review we have described general relativity in terms
of the tetrad fields $e_{a\mu}$ and of the torsion tensor 
$T_{a\mu\nu}=\partial_\mu e_{a\nu}-\partial_\nu e_{a\mu}$. The tetrad fields
constitute the frame adapted to observers in space-time. All observers are
allowed, and to each one there is a frame adapted to its worldline. 
This alternative description does not imply an alternative dynamics for
the metric tensor. The tetrad fields satisfy field equations that are 
strictly equivalent to Einstein's equations. In this geometrical
description, the tetrad fields yield several new definitions that cannot be
established in the ordinary metrical formulation. The field
equations lead to an actual conservation equation, and to consistent 
definitions of the energy, momentum and angular momentum of the gravitational
field. In the analysis of some standard configurations of the gravitational
field, these definitions lead to results that are consistent with the physical
configuration. The definitions are not invariant or covariant under local 
SO(3,1) transformations, but only covariant under global transformations. 
Invariance of field quantities under local SO(3,1) transformations imply 
that the measurement of these quantities is the same in inertial and
accelerated frames. This invariance is not a natural feature of concepts such 
as energy, momentum and angular momentum. Energy
is always the zero component of an energy-momentum vector.

Although teleparallel gravity was first addressed by Hayashi and Shirafuji
\cite{Hay} in a geometrical framework similar to the one adopted here, it
may be considered as a limiting case of the more general framework of
metric-affine theories of gravity. In this context, the gravitational field is
described both by the tetrad fields and an independent affine connection, and 
the theory exhibits explicit invariance under local SO(3,1) 
transformations. However, one has to deal with Lagrange multipliers that 
enforce the vanishing of the curvature tensor of the connection, 
and one also has field equations for the zero-curvature connection 
\cite{Hehl,Nitsch,Maluf1,OP}. The geometrical framework is more intricate, 
and it is not clear that the initial value problem is well established for 
all field quantities. There is an ambiguity in the 
determination of the Lagrange multipliers \cite{OP}. Moreover, in our opinion 
local Lorentz invariance is not a natural feature of distant parallelism, or
teleparallelism.

The TEGR is geometrically different from Einstein-Cartan
type theories. The latter are theories with both metric and torsion as
independent field quantities, and the torsion may or may not propagate in
space-time. In these theories, torsion is an additional geometrical entity 
related to spinning matter. In the TEGR, torsion plays a relevant role in
both the kinematic and dynamical description of the gravitational field,
as we have seen.

Since the TEGR is formulated in terms of tetrad fields, one may construct
the space-time curvature tensor. Of course the curvature tensor is of utmost
importance in the ordinary metrical formulation of
general relativity. The curvature tensor is non-vanishing in general, but it
does not play a major role in the formulation of the TEGR.
We argued in the Introduction that a theory formulated in terms of tetrad 
fields is geometrically more rich than the metrical formulation, because
one may dispose of the concepts of both the Weitzenb\"{o}ck and Riemannian 
geometries. The torsion tensor depends on first order derivatives of the
tetrad fields, and is geometrically simpler than the curvature tensor, which
depends on second order derivatives.

We have presented the Hamiltonian formulation and the constraint algebra of 
the theory. All constraints are first class constraints. Therefore, the time 
evolution of all field quantities is well defined. As a consequence of the 
Hamiltonian formulation, the initial value problem in the realm of the TEGR 
is mathematically and physically consistent.

In summary, the TEGR is a simple and consistent description of the 
gravitational field. It embodies all physical features of the standard 
metrical formulation, and allows definitions for the energy, momentum and
angular momentum of the gravitational field that satisfy the algebra of the
Poincar\'e group in the phase space of the theory.\par

\bigskip

\noindent {\bf Acknowledgement}\par
\noindent 
The author is grateful to J. F. da Rocha-Neto, S. C. Ulhoa and F. F. Faria,
for the participation and essential contribution in the papers that
allowed to prepare this review, and to S. C. Ulhoa also for a careful reading
of the manuscript.

\end{document}